\documentclass[12pt]{article}
\begin{document}

\author{E. M. Cioroianu\thanks{%
E-mail address: manache@central.ucv.ro} and S. C. S\u {a}raru\thanks{%
E-mail address: scsararu@central.ucv.ro} \\
Faculty of Physics, University of Craiova,\\
13 A. I. Cuza Str., Craiova 200585, Romania}
\title{Two-dimensional interactions between a BF-type theory and a collection of
vector fields }
\maketitle

\begin{abstract}
Consistent interactions that can be added to a two-dimensional, free abelian
gauge theory comprising a special class of BF-type models and a collection
of vector fields are constructed from the deformation of the solution to the
master equation based on specific cohomological techniques. The deformation
procedure modifies the Lagrangian action, the gauge transformations, as well
as the accompanying algebra of the interacting model.

\end{abstract}

\section{Introduction}

A key point in the development of the BRST formalism was its cohomological
understanding, which allowed, among others, a useful investigation of many
interesting aspects related to the perturbative renormalization problem~\cite
{4}--\cite{5}, the anomaly-tracking mechanism~\cite{5}--\cite{6}, the
simultaneous study of local and rigid invariances of a given theory~\cite{7}%
, as well as to the reformulation of the construction of consistent
interactions in gauge theories~\cite{7a}--\cite{7d} in terms of the
deformation theory~\cite{8a}--\cite{8c}, or, actually, in terms of the
deformation of the solution to the master equation.

The scope of this paper is to investigate the consistent interactions that
can be added to a free, abelian, two-dimensional gauge theory consisting of
a collection of vector fields and a BF-type model~\cite{13} involving a set
of scalar fields, two collections of one-forms and a system of two-forms.
This work enhances the previous Lagrangian~\cite{mpla} and Hamiltonian~\cite
{ijmpa}--\cite{jhep} results on the study of self-interactions in certain
classes of BF-type models. The resulting interactions are accurately
described by a gauge theory with an open algebra of gauge transformations.
The interacting model reveals a geometric interpretation in terms of a
Poisson structure present in various models of two-dimensional gravity~\cite
{grav1}--\cite{grav4} and also some interesting algebraic features. The
analysis of Poisson Sigma Models, including their relationship to
two-dimensional gravity and the study of classical solutions, can be found
in~\cite{psm1}--\cite{psmn} (see also~\cite{ikeda12}).

Our strategy goes as follows. Initially, we determine in Section 2 the
antifield-BRST symmetry of the free model, that splits as the sum between
the Koszul-Tate differential and the exterior derivative along the gauge
orbits, $s=\delta +\gamma $. Next, in Section 3 we determine the consistent
deformations of the solution to the master equation for the free model. The
first-order deformation belongs to the local cohomology $H^{0}(s|d)$, where $%
d$ is the exterior space-time derivative. The computation of the
cohomological space $H^{0}(s|d)$ proceeds by expanding the co-cycles
according to the antighost number, and by further using the cohomological
groups $H(\gamma )$ and $H(\delta |d)$. We find that the first-order
deformation is parametrized by some functions of the undifferentiated scalar
fields, which become restricted to fulfill certain equations in order to
produce a consistent second-order deformation. With the help of these
equations, we then infer that the remaining deformations, of order three and
higher, can be taken to vanish. The identification of the interacting model
is developed in Section 4. The cross-couplings between the collection of
vector fields and the BF field spectrum are described, among others, by
generalized cubic and quartic Yang-Mills vertices in some ``backgrounds'' of
scalar fields. Meanwhile, both the gauge transformations corresponding to
the coupled model and their algebra are deformed with respect to the initial
abelian theory in such a way that the new gauge algebra becomes open.
Section 5 comments on two classes of solutions to the equations satisfied by
the various functions of the scalar fields (that parametrize the deformed
solution to the master equation) and Section 6 closes the paper with the
main conclusions.

\section{Free model. Antibracket-antifield BRST symmetry}

The starting point is represented by the free Lagrangian action 
\begin{equation}
S_{0}\left[ A_{\mu }^{a},H_{\mu }^{a},\varphi _{a},B_{a}^{\mu \nu },V_{\mu
}^{A}\right] =\int d^{2}x\left( H_{\mu }^{a}\partial ^{\mu }\varphi _{a}+%
\frac{1}{2}B_{a}^{\mu \nu }F_{\mu \nu }^{\prime a}-\frac{1}{4}F_{\mu \nu
}^{A}F_{A}^{\mu \nu }\right) ,  \label{f1}
\end{equation}
where we used the notations 
\begin{equation}
F_{\mu \nu }^{A}=\partial _{[\mu }V_{\nu ]}^{A},\;F_{\mu \nu }^{\prime
a}=\partial _{[\mu }A_{\nu ]}^{a},  \label{f1a}
\end{equation}
and the symbol $\left[ \mu \nu \cdots \right] $ denotes full
antisymmetrization with respect to the indices between brackets, but without
normalization factors. Capital Latin indices $A$, $B$, etc., are raised with
a constant, symmetric and field-independent, non-degenerate matrix $k^{AB}$.
We observe that~(\ref{f1}) is written like a sum between the action of a
two-dimensional abelian BF theory (involving two sets of one-forms $\left\{
A_{\mu }^{a},H_{\mu }^{a}\right\} $, a collection of scalar fields $\left\{
\varphi _{a}\right\} $ and a sequence of two forms $\left\{ B_{a}^{\mu \nu
}\right\} $) and the action corresponding to a set of abelian vector fields $%
\left\{ V_{\mu }^{A}\right\} $. The collection indices $a$, $b$, etc., and
respectively $A$, $B$, etc., are assumed to run independently ones to the
others. A generating set of gauge transformations for the action~(\ref{f1})
can be taken under the form 
\begin{eqnarray}
\delta _{\epsilon }A_{\mu }^{a} &=&\partial _{\mu }\epsilon ^{a},\;\delta
_{\epsilon }H_{\mu }^{a}=\partial ^{\nu }\epsilon _{\mu \nu }^{a},
\label{f2} \\
\delta _{\epsilon }\varphi _{a} &=&0=\delta _{\epsilon }B_{a}^{\mu \nu
},\;\delta _{\epsilon }V_{\mu }^{A}=\partial _{\mu }\epsilon ^{A},
\label{f3}
\end{eqnarray}
where all the gauge parameters are bosonic, with $\epsilon _{\mu \nu }^{a}$
antisymmetric in their Lorentz indices. In $D=2$ spacetime dimensions, where
this model evolves, the abelian gauge transformations~(\ref{f2}--\ref{f3})
are irreducible. In conclusion,~(\ref{f1}) and~(\ref{f2}--\ref{f3}) describe
a linear (the field equations are linear in the fields) gauge theory of
Cauchy order equal to two.

In order to construct the antifield-BRST symmetry for the free gauge theory
under study, we need to identify the algebra on which the BRST differential
acts. The generators of the BRST algebra are, besides the original bosonic
fields 
\begin{equation}
\Phi ^{\alpha _{0}}=\left( A_{\mu }^{a},H_{\mu }^{a},\varphi _{a},B_{a}^{\mu
\nu },V_{\mu }^{A}\right) ,  \label{f10a}
\end{equation}
the fermionic ghosts 
\begin{equation}
\eta ^{\alpha _{1}}=\left( \eta ^{a},\eta _{\mu \nu }^{a},C^{A}\right) ,
\label{f10b}
\end{equation}
respectively associated with the gauge parameters from~(\ref{f2}--\ref{f3}),
as well as their antifields 
\begin{equation}
\Phi _{\alpha _{0}}^{*}=\left( A_{a}^{*\mu },H_{a}^{*\mu },\varphi
^{*a},V_{A}^{*\mu },B_{\mu \nu }^{*a}\right) ,\;\eta _{\alpha
_{1}}^{*}=\left( \eta _{a}^{*},\eta _{a}^{*\mu \nu },C_{A}^{*}\right) .
\label{f11}
\end{equation}
The Grassmann parity of an antifield is opposite to that of the
corresponding field/ ghost. It is understood that $\eta _{\mu \nu }^{a}$ and 
$\eta _{a}^{*\mu \nu }$ are antisymmetric, just like the gauge parameters $%
\epsilon _{\mu \nu }^{a}$. Since the gauge generators of this model are
field-independent, it follows that the BRST differential $s$ simply reduces
to 
\begin{equation}
s=\delta +\gamma ,  \label{f11a}
\end{equation}
where $\delta $ represents the Koszul-Tate differential, graded by the
antighost number $\mathrm{agh}$ ($\mathrm{agh}\left( \delta \right) =-1$),
and $\gamma $ stands for the exterior derivative along the gauge orbits,
whose degree is named pure ghost number $\mathrm{pgh}$ ($\mathrm{pgh}\left(
\gamma \right) =1$). Naturally, these two degrees do not interfere ($\mathrm{%
agh}\left( \gamma \right) =0$, $\mathrm{pgh}\left( \delta \right) =0$). The
overall degree that grades the BRST complex, known as the ghost number ($%
\mathrm{gh}$), is defined like the difference between the pure ghost number
and the antighost number, such that $\mathrm{gh}\left( s\right) =\mathrm{gh}%
\left( \delta \right) =\mathrm{gh}\left( \gamma \right) =1$. According to
the standard rules of the BRST method, the corresponding degrees of the
generators from the BRST complex are valued like 
\begin{eqnarray}
\mathrm{pgh}\left( \Phi ^{\alpha _{0}}\right) &=&\mathrm{pgh}\left( \Phi
_{\alpha _{0}}^{*}\right) =0,\;\mathrm{pgh}\left( \eta ^{\alpha _{1}}\right)
=1,\;\mathrm{pgh}\left( \eta _{\alpha _{1}}^{*}\right) =0,  \label{f8} \\
\mathrm{agh}\left( \Phi ^{\alpha _{0}}\right) &=&0,\;\mathrm{agh}\left( \Phi
_{\alpha _{0}}^{*}\right) =1,\;\mathrm{agh}\left( \eta ^{\alpha _{1}}\right)
=0,\;\mathrm{agh}\left( \eta _{\alpha _{1}}^{*}\right) =2.  \label{f9}
\end{eqnarray}
Actually,~(\ref{f11a}) is a decomposition of the BRST differential according
to the antighost number and it shows that $s$ contains only components of
antighost number equal to minus one and zero. Consequently, the equation
expressing the nilpotency of $s$ projected on the distinct values of the
antighost number is equivalent with the nilpotency and anticommutation of
its components 
\begin{equation}
s^{2}=0\Leftrightarrow \left( \delta ^{2}=0,\;\delta \gamma +\gamma \delta
=0,\;\gamma ^{2}=0\right) .  \label{f11b}
\end{equation}
The Koszul-Tate differential is imposed to realize an homological resolution
of the algebra of smooth functions defined on the stationary surface of the
field equations for the action~(\ref{f1}), while the exterior longitudinal
derivative is related to the gauge symmetries~(\ref{f2}--\ref{f3}) through
its cohomology at pure ghost number zero computed in the cohomology of $%
\delta $, which is required to be the algebra of physical observables for
the free model under consideration. The actions of $\delta $ and $\gamma $
on the generators from the BRST complex, which enforce all the above
mentioned properties, are given by 
\begin{equation}
\delta \Phi ^{\alpha _{0}}=0=\delta \eta ^{\alpha _{1}},  \label{f12}
\end{equation}
\begin{equation}
\delta A_{a}^{*\mu }=\partial _{\alpha }B_{a}^{\alpha \mu },\;\delta
H_{a}^{*\mu }=-\partial ^{\mu }\varphi _{a},\;\delta \varphi ^{*a}=\partial
^{\mu }H_{\mu }^{a},\;\delta B_{\mu \nu }^{*a}=-\frac{1}{2}F_{\mu \nu
}^{\prime a},  \label{f13}
\end{equation}
\begin{equation}
\delta V_{A}^{*\mu }=-\partial _{\alpha }F_{A}^{\alpha \mu },\;\delta \eta
_{a}^{*}=-\partial _{\mu }A_{a}^{*\mu },\;\delta \eta _{a}^{*\mu \nu }=\frac{%
1}{2}\partial ^{[\mu }H_{a}^{*\nu ]},\;\delta C_{A}^{*}=-\partial _{\mu
}V_{A}^{*\mu },  \label{f14}
\end{equation}
\begin{equation}
\gamma \Phi _{\alpha _{0}}^{*}=0=\gamma \eta _{\alpha _{1}}^{*},\;\gamma
V_{\mu }^{A}=\partial _{\mu }C^{A},  \label{f15}
\end{equation}
\begin{equation}
\gamma A_{\mu }^{a}=\partial _{\mu }\eta ^{a},\;\gamma H_{\mu }^{a}=\partial
^{\nu }\eta _{\mu \nu }^{a},\;\gamma \varphi _{a}=\gamma B_{a}^{\mu \nu
}=\gamma \eta ^{\alpha _{1}}=0.  \label{f16}
\end{equation}

The Lagrangian BRST differential admits a canonical action in a structure
named antibracket and defined by decreeing the fields/ghosts conjugated with
the corresponding antifields, $s\cdot =\left( \cdot ,S\right) $, where $%
\left( ,\right) $ signifies the antibracket and $S$ denotes the canonical
generator of the BRST symmetry. It is a bosonic functional of ghost number
zero, involving both field/ghost and antifield spectra, that obeys the
master equation 
\begin{equation}
\left( S,S\right) =0.  \label{mastorig}
\end{equation}
The master equation is equivalent with the second-order nilpotency of $s$,
where its solution $S$ encodes the entire gauge structure of the associated
theory. Taking into account the formulas~(\ref{f12}--\ref{f16}), as well as
the standard actions of $\delta $ and $\gamma $ in canonical form, we find
that the complete solution to the master equation for the model under study
reads 
\begin{eqnarray}
S &=&S_{0}\left[ A_{\mu }^{a},H_{\mu }^{a},\varphi _{a},B_{a}^{\mu \nu
},V_{\mu }^{A}\right]  \nonumber \\
&&+\int d^{2}x\left( A_{a}^{*\mu }\partial _{\mu }\eta ^{a}+H_{a}^{*\mu
}\partial ^{\nu }\eta _{\mu \nu }^{a}+V_{A}^{*\mu }\partial _{\mu
}C^{A}\right) ,  \label{f5}
\end{eqnarray}
such that it contains pieces of antighost number zero and one. The absence
of components with antighost numbers higher than one is due to the
abelianity and irreducibility of the chosen generating set of gauge
transformations. If the gauge algebra were non-abelian, then the solution to
the master equation would also include terms of antighost number two that
are quadratic in the ghosts~(\ref{f10b}): ones linear in the antifields $%
\eta _{\alpha _{1}}^{*}$ and proportional with the structure functions
appearing at the commutators between the gauge generators, and others
quadratic in the antifields $\Phi _{\alpha _{0}}^{*}$. The latter kind of
elements is present only if these commutators strictly close on-shell, \textit{i.e.},
on the stationary surface of field equations, or, which is the same, only if
the gauge algebra is open. In the case where the gauge algebra is open, the
solution to the master equation may in principle continue to be
non-vanishing at antighost numbers higher than two, the corresponding
elements being related to the higher-order structure functions and to the
identities satisfied by them.

The main ingredients of the antifield-BRST symmetry derived in this section
will be useful in the sequel at the analysis of consistent interactions that
can be added to the action~(\ref{f1}) without changing its number of
independent gauge symmetries.

\section{Deformation of the solution to the master equation}

\subsection{General idea}

A consistent deformation of the free action~(\ref{f1}) and of its gauge
invariances~(\ref{f2}--\ref{f3}) defines a deformation of the corresponding
solution to the master equation that preserves both the master equation and
the field/ antifield spectra. Let us denote by $g$ the coupling constant and
assume that the local functional $\bar{S}_{0}\left[ A_{\mu }^{a},H_{\mu
}^{a},\varphi _{a},B_{a}^{\mu \nu },V_{\mu }^{A}\right] =S_{0}+g\int
d^{2}xa_{0}+g^{2}\int d^{2}xb_{0}+O\left( g^{3}\right) $ represents a
consistent deformation of~(\ref{f1}), subject to the deformed gauge
transformations $\bar{\delta}_{\epsilon }A_{\mu }^{a}=\partial _{\mu
}\epsilon ^{a}+g\lambda _{\mu }^{a}+g^{2}\lambda _{\mu }^{\prime a}+O\left(
g^{3}\right) $, $\bar{\delta}_{\epsilon }H_{\mu }^{a}=\partial ^{\nu
}\epsilon _{\mu \nu }^{a}+g\rho _{\mu }^{a}+g^{2}\rho _{\mu }^{\prime
a}+O\left( g^{3}\right) $, $\bar{\delta}_{\epsilon }\varphi _{a}=g\sigma
_{a}+g^{2}\sigma _{a}^{\prime }+O\left( g^{3}\right) $, $\bar{\delta}%
_{\epsilon }B_{a}^{\mu \nu }=g\sigma _{a}^{\mu \nu }+g^{2}\sigma
_{a}^{\prime \mu \nu }+O\left( g^{3}\right) $ and $\bar{\delta}_{\epsilon
}V_{\mu }^{A}=\partial _{\mu }\epsilon ^{A}+gv_{\mu }^{A}+g^{2}v_{\mu
}^{\prime A}+O\left( g^{3}\right) $ (by consistent we mean that $\bar{S}_{0}$
is invariant under the modified gauge transformations $\bar{\delta}%
_{\epsilon }\Phi ^{\alpha _{0}}$ at all orders in the coupling constant).
Accordingly, we find that 
\begin{equation}
\bar{S}=S+g\int d^{2}xa+g^{2}\int d^{2}xb+O\left( g^{3}\right) ,  \label{f4}
\end{equation}
is a consistent deformed solution to the master equation for the interacting
theory, \textit{i.e.}, it satisfies the equation 
\begin{equation}
\left( \bar{S},\bar{S}\right) =0,  \label{mastdef}
\end{equation}
at all orders in the coupling constant (with $S$ given by~(\ref{f5})).
Moreover, the objects $a$ and $b$ start like 
\begin{eqnarray}
a &=&a_{0}+A_{a}^{*\mu }\bar{\lambda}_{\mu }^{a}+H_{a}^{*\mu }\bar{\rho}%
_{\mu }^{a}+V_{A}^{*\mu }\bar{v}_{\mu }^{A}+\varphi ^{*a}\bar{\sigma}%
_{a}+B_{\mu \nu }^{*a}\bar{\sigma}_{a}^{\mu \nu }+``\mathrm{more}",
\label{f6} \\
b &=&b_{0}+A_{a}^{*\mu }\bar{\lambda}_{\mu }^{\prime a}+H_{a}^{*\mu }\bar{%
\rho}_{\mu }^{\prime a}+V_{A}^{*\mu }\bar{v}_{\mu }^{\prime A}+\varphi ^{*a}%
\bar{\sigma}_{a}^{\prime }+B_{\mu \nu }^{*a}\bar{\sigma}_{a}^{\prime \mu \nu
}+``\mathrm{more}",  \label{f7}
\end{eqnarray}
where the ``bar'' quantities are obtained by replacing the gauge parameters $%
\epsilon ^{a}$, $\epsilon _{\mu \nu }^{a}$ and $\epsilon ^{A}$ respectively
with the fermionic ghosts $\eta ^{a}$, $\eta _{\mu \nu }^{a}$ and $C^{A}$ in
the functions $\lambda _{\mu }^{a}$, $\lambda _{\mu }^{\prime a}$, $\rho
_{\mu }^{a}$, $\rho _{\mu }^{\prime a}$, $\sigma _{a}$, $\sigma _{a}^{\prime
}$, $\sigma _{a}^{\mu \nu }$, $\sigma _{a}^{\prime \mu \nu }$, $v_{\mu }^{A}$
and $v_{\mu }^{\prime A}$ contained in the deformed gauge transformations.

\subsection{First-order deformation}

\subsubsection{Basic cohomological results}

Using the development~(\ref{f4}) and the Eq.~(\ref{mastorig}) satisfied
by $S$, we obtain that the master equation~(\ref{mastdef}) of the deformed
theory holds to order $g$ if and only if 
\begin{equation}
sa=\partial _{\mu }j^{\mu },  \label{f17}
\end{equation}
for some local $j^{\mu }$. This means that the non-integrated density of the
first-order deformation of the solution to the master equation, $a$, belongs
to the local cohomology of the BRST differential, $H^{0}\left( s|d\right) $,
where $d$ is the exterior spacetime derivative. In the case where $a$ is a $s
$ coboundary modulo $d$ ($a=sc+\partial _{\mu }k^{\mu }$), then the
deformation is trivial (it can be eliminated by a redefinition of the
fields). As a consequence, $a$ is unique only up to replacing it with an
element from the same cohomological class, $a\rightarrow a+sc+\partial _{\mu
}k^{\mu }$, and, on the other hand, if $a$ is purely trivial, $a=sc+\partial
_{\mu }k^{\mu }$, then it can be removed from $\bar{S}$ by setting $a=0$.
For obvious reasons, we are interested only in smooth, local,
Lorentz-covariant and Poincar\'{e}-invariant deformations. In order to
investigate the solution to~(\ref{f17}), we develop $a$ according to the
antighost number 
\begin{equation}
a=a_{0}+a_{1}+\cdots +a_{J},\;\mathrm{agh}\left( a_{k}\right) =k=\mathrm{pgh}%
\left( a_{k}\right) ,\;\varepsilon \left( a_{k}\right) =0,  \label{f18}
\end{equation}
and assume that the expansion~(\ref{f18}) stops at a finite, but otherwise
arbitrary, value of the antighost number, $J$. (The notation $\varepsilon
\left( F\right) $ signifies the Grassmann parity of $F$.) This result can be
argued like in ~\cite{gen2} (Section 3), under the sole assumption that the
interacting Lagrangian at the first order in the coupling constant, $a_{0}$,
is local, so it contains a finite, but otherwise arbitrary, number of
derivatives. Replacing~(\ref{f18}) into the Eq.~(\ref{f17}) and taking
into account the decomposition~(\ref{f11a}), we obtain that the Eq.~(%
\ref{f17}) is equivalent to a tower of local equations, corresponding to the
different decreasing values of the antighost number 
\begin{eqnarray}
\gamma a_{J} &=&\partial _{\mu }j_{J}^{\mu },  \label{f18b} \\
\delta a_{J}+\gamma a_{J-1} &=&\partial _{\mu }j_{J-1}^{\mu },  \label{f18c}
\\
\delta a_{k}+\gamma a_{k-1} &=&\partial _{\mu }j_{k-1}^{\mu },\;J-1\geq
k\geq 1,  \label{f18d}
\end{eqnarray}
where $\left( j_{k}^{\mu }\right) _{k=\overline{0,J}}$ are some local
currents with $\mathrm{agh}\left( j_{k}^{\mu }\right) =k$. As $\mathrm{pgh}%
\left( a_{J}\right) =J$, the Eq.~(\ref{f18b}) shows that $a_{J}$
belongs to the local cohomology of the exterior derivative along the gauge
orbits at pure ghost number $J$, $H^{J}\left( \gamma |d\right) $. Following
a reasoning similar to that from~\cite{gen2}--\cite{lingr}, it can be shown
that one can replace the Eq.~(\ref{f18b}) at strictly positive
antighost numbers with 
\begin{equation}
\gamma a_{J}=0,\;J>0.  \label{f18a}
\end{equation}
In other words, for $J>0$ the last representative from~(\ref{f18}) can
always be considered to pertain to the cohomological group of the exterior
derivative along the gauge orbits at pure ghost number $J$, $H^{J}\left(
\gamma \right) $. As a consequence, it is unique up to $\gamma $-exact
contributions, $a_{J}\rightarrow a_{J}+\gamma c_{J}$, while the purely $%
\gamma $-trivial solutions $a_{J}=\gamma c_{J}$ can be safely removed by
taking $a_{J}=0$. In conclusion, the Eq.~(\ref{f17}) associated with
the local form of the first-order deformation is completely equivalent to
the tower of Eqs.~(\ref{f18a}) and~(\ref{f18c}--\ref{f18d}).

Thus, we need to know the cohomology of $\gamma $, $H\left( \gamma \right) $%
, in order to determine the terms of highest antighost number in $a$. From
the definitions~(\ref{f15}--\ref{f16}) it is simple to see that this
cohomology is generated by $F_{\mu \nu }^{A}$, $F_{\mu \nu }^{\prime a}$, $%
\varphi _{a}$, $B_{a}^{\mu \nu }$, $\partial ^{\mu }H_{\mu }^{a}$, by the
antifields $\left\{ \Phi _{\alpha _{0}}^{*}\right\} $, by all their
derivatives, as well as by the undifferentiated ghosts $\eta ^{\alpha _{1}}$%
. (The derivatives of the ghosts are $\gamma $-exact, as can be observed
from the last relation in~(\ref{f15}) and the first two formulas in~(\ref
{f16}), so we can discard them as being trivial in $H\left( \gamma \right) $%
.) If we denote by $e^{M}\left( \eta ^{\alpha _{1}}\right) $ the elements of
pure ghost number equal to $M$ of a basis in the space of polynomials in $%
\eta ^{\alpha _{1}}$, which is finite-dimensional due to the anticommutation
of the ghosts, it follows that the general local solution to the Eq.~(%
\ref{f18a}) takes the form (up to irrelevant, $\gamma $-exact contributions) 
\begin{equation}
a_{J}=\alpha _{J}\left( \left[ F_{\mu \nu }^{A}\right] ,\left[ F_{\mu \nu
}^{\prime a}\right] ,\left[ \varphi _{a}\right] ,\left[ B_{a}^{\mu \nu
}\right] ,\left[ \partial ^{\mu }H_{\mu }^{a}\right] ,\left[ \Phi _{\alpha
_{0}}^{*}\right] ,\left[ \eta _{\alpha _{1}}^{*}\right] \right) e^{J}\left(
\eta ^{\alpha _{1}}\right) ,\;J>0,  \label{f19}
\end{equation}
where $\mathrm{agh}\left( \alpha _{J}\right) =J$ for $a_{J}$ to have the
ghost number equal to zero, and $\alpha _{J}$ must display the same
Grassmann parity like $e^{J}$ in order to ensure that $a_{J}$ is bosonic.
Here and in the sequel the notation $f\left( \left[ q\right] \right) $
signifies that $f$ depends on $q$ and its spacetime derivatives up to a
finite order. The index-notation $J$ is generic, in the sense that it may
include unspecified Lorentz and/or collection indices. As they have both
finite antighost number and derivative order, the elements $\alpha _{J}$,
which are non-trivial in $H^{0}\left( \gamma \right) $, are polynomials in
the antifields, their derivatives, and in the allowed derivatives of the
fields, but may contain an indefinite number of undifferentiated fields $%
\varphi _{a}$ and $B_{a}^{\mu \nu }$. They will be called ``invariant
polynomials''. At zero antighost number, the invariant polynomials are
nothing but the local, gauge-invariant quantities of the free model under
study. The fact that we can replace the Eq.~(\ref{f18b}) for $J>0$ with
(\ref{f18a}) is a consequence of the triviality of the cohomology of the
exterior spacetime differential in the space of invariant polynomials at
strictly positive antighost numbers. So, if $\alpha _{J}$ is an invariant
polynomial with $\mathrm{agh}\left( \alpha _{J}\right) =J>0$ that is $d$%
-closed, $d\alpha _{J}=0$, then $\alpha _{J}=d\beta _{J}$, with $\beta _{J}$
also an invariant polynomial.

Inserting the expression~(\ref{f19}) into the Eq.~(\ref{f18c}) and
recalling the definitions~(\ref{f15}--\ref{f16}), we find that a necessary
condition for the existence of (non-trivial) solutions $a_{J-1}$ is that the
invariant polynomials $\alpha _{J}$ are (non-trivial) elements from the
local cohomology group of the Koszul-Tate differential at pure ghost number
zero and at strictly positive antighost number $J$, $H_{J}\left( \delta
|d\right) $\footnote{%
We note that the local cohomology group of the Koszul-Tate differential at
strictly positive pure ghost \textit{and} antighost numbers is trivial, so
the notations $H_{J}\left( \delta |d\right) $ and $H\left( \delta |d\right) $
automatically take into consideration only objects of pure ghost number zero
(see, for instance,~\cite{gen1} and~\cite{commun1}).} 
\begin{equation}
\delta \alpha _{J}=\partial _{\mu }k_{J-1}^{\mu },\;\mathrm{agh}\left(
k_{J-1}^{\mu }\right) =J-1\geq 0.  \label{f19a}
\end{equation}
By ``trivial elements of $H_{J}\left( \delta |d\right) $'' we understand $%
\delta $-exact modulo $d$ objects, hence of the form $\delta
d_{J+1}+\partial _{\mu }m_{J}^{\mu }$. As a consequence of~(\ref{f19a}), we
need to investigate some of the main properties of the cohomology $H\left(
\delta |d\right) $ at strictly positive antighost numbers in order to fully
determine the component $a_{J}$ of highest antighost number from the
first-order deformation. As we have discussed in Section 2, the free model
under study is a linear gauge theory of Cauchy order equal to two. In
agreement with the general results from~\cite{gen1} (also see~\cite{gen2}--%
\cite{lingr}), one can state that $H\left( \delta |d\right) $ (at pure ghost
number zero) is trivial at antighost numbers strictly greater than its
Cauchy order. The same result holds for the local cohomology of the
Koszul-Tate differential in the space of invariant polynomials, $H^{\mathrm{%
inv}}\left( \delta |d\right) $, so we actually have that 
\begin{equation}
H_{J}\left( \delta |d\right) =0,\;H_{J}^{\mathrm{inv}}\left( \delta
|d\right) =0,\;J>2.  \label{f19b}
\end{equation}
An element of $H_{J}^{\mathrm{inv}}\left( \delta |d\right) $ is defined via
an equation similar to~(\ref{f19a}), but with the corresponding current $%
k_{J-1}^{\mu }$ an invariant polynomial. Moreover, it can be shown~\cite
{gen2}--\cite{lingr} that if the invariant polynomial $\alpha _{J}$ with $%
\mathrm{agh}\left( \alpha _{J}\right) =J\geq 2$ is trivial in $H_{J}\left(
\delta |d\right) $, then it can be taken to be trivial also in $H_{J}^{%
\mathrm{inv}}\left( \delta |d\right) $, \textit{i.e.}, 
\begin{equation}
\left( \alpha _{J}=\delta d_{J+1}+\partial _{\mu }m_{J}^{\mu },\;\mathrm{agh}%
\left( \alpha _{J}\right) =I\geq 2\right) \Rightarrow \alpha _{J}=\delta
\beta _{J+1}+\partial _{\mu }\gamma _{J}^{\mu },  \label{f19c}
\end{equation}
with both $\beta _{J+1}$ and $\gamma _{J}^{\mu }$ invariant polynomials.
With the help of the definitions~(\ref{f12}--\ref{f14}), we find that the
most general non-trivial representative from $H_{2}\left( \delta |d\right) $%
, which, essentially, has the same status in $H_{2}^{\mathrm{inv}}\left(
\delta |d\right) $, is 
\begin{equation}
\alpha _{2}^{0}=K^{\Delta }M_{\Delta }+K^{\Delta ^{\prime }\mu \nu
}N_{\Delta ^{\prime }\mu \nu }+K^{\Delta ^{\prime \prime }}P_{\Delta
^{\prime \prime }},  \label{f20}
\end{equation}
where $M_{\Delta }$, $N_{\Delta ^{\prime }\mu \nu }$ and $P_{\Delta ^{\prime
\prime }}$ are respectively given by 
\begin{eqnarray}
M_{\Delta } &=&\frac{\partial W_{\Delta }}{\partial \varphi _{m}}\eta
_{m}^{*}-\frac{\partial ^{2}W_{\Delta }}{\partial \varphi _{m}\partial
\varphi _{n}}\left( B_{m}^{\mu \nu }\eta _{n\mu \nu }^{*}+H_{m\mu
}^{*}A_{n}^{*\mu }\right)   \nonumber \\
&&-\frac{1}{2}\frac{\partial ^{3}W_{\Delta }}{\partial \varphi _{m}\partial
\varphi _{n}\partial \varphi _{p}}H_{m\mu }^{*}H_{n\nu }^{*}B_{p}^{\mu \nu },
\label{f21}
\end{eqnarray}
\begin{equation}
N_{\Delta ^{\prime }\mu \nu }=\frac{\partial U_{\Delta ^{\prime }}}{\partial
\varphi _{m}}\eta _{m\mu \nu }^{*}+\frac{1}{2}\frac{\partial ^{2}U_{\Delta
^{\prime }}}{\partial \varphi _{m}\partial \varphi _{n}}H_{m\mu }^{*}H_{n\nu
}^{*},  \label{f22}
\end{equation}
\begin{equation}
P_{\Delta ^{\prime \prime }}=f_{\Delta ^{\prime \prime }}^{A}C_{A}^{*}+\frac{%
\partial f_{\Delta ^{\prime \prime }}^{A}}{\partial \varphi _{m}}\left(
V_{A}^{*\mu }H_{m\mu }^{*}+\eta _{m\mu \nu }^{*}F_{A}^{\mu \nu }\right) +%
\frac{1}{2}\frac{\partial ^{2}f_{\Delta ^{\prime \prime }}^{A}}{\partial
\varphi _{m}\partial \varphi _{n}}H_{m\mu }^{*}H_{n\nu }^{*}F_{A}^{\mu \nu },
\label{f23}
\end{equation}
and $K^{\Delta }$, $K^{\Delta ^{\prime }\mu \nu }$ and $K^{\Delta ^{\prime
\prime }}$ denote some constant coefficients, with $K^{\Delta ^{\prime }\mu
\nu }$ antisymmetric in their Lorentz indices. The generic indices $\Delta $%
, $\Delta ^{\prime }$ and $\Delta ^{\prime \prime }$ are exclusively
composed of collection indices (of the type $a$, $b$, etc., and/or $A$, $B$,
etc.). All the functions $W_{\Delta }$, $U_{\Delta ^{\prime }}$ and $%
f_{\Delta ^{\prime \prime }}^{A}$ involved in~(\ref{f21}--\ref{f23}) depend
in an arbitrary manner on the undifferentiated scalar fields $\varphi _{a}$.
Moreover, the objects $M_{\Delta }$, $N_{\Delta ^{\prime }\mu \nu }$ and $%
P_{\Delta ^{\prime \prime }}$ separately satisfy the equations 
\begin{equation}
\delta M_{\Delta }=\partial _{\mu }k_{\Delta }^{\mu },\;\delta N_{\Delta
^{\prime }\mu \nu }=\frac{1}{2}\partial _{\left[ \mu \right. }k_{\Delta
^{\prime }\left. \nu \right] },\;\delta P_{\Delta ^{\prime \prime
}}=\partial _{\mu }k_{\Delta ^{\prime \prime }}^{\mu },  \label{f24}
\end{equation}
where the corresponding currents are also invariant polynomials 
\begin{eqnarray}
k_{\Delta }^{\mu } &=&-\left( \frac{\partial W_{\Delta }}{\partial \varphi
_{m}}A_{m}^{*\mu }+\frac{\partial ^{2}W_{\Delta }}{\partial \varphi
_{m}\partial \varphi _{n}}B_{m}^{\mu \nu }H_{n\nu }^{*}\right) ,
\label{f24a} \\
k_{\Delta ^{\prime }\nu } &=&\frac{\partial U_{\Delta ^{\prime }}}{\partial
\varphi _{m}}H_{m\nu }^{*},  \label{f24b} \\
k_{\Delta ^{\prime \prime }}^{\mu } &=&-V_{A}^{*\mu }f_{\Delta ^{\prime
\prime }}^{A}+\frac{\partial f_{\Delta ^{\prime \prime }}^{A}}{\partial
\varphi _{m}}F_{A}^{\mu \nu }H_{m\nu }^{*},  \label{f24c}
\end{eqnarray}
and hence we have that 
\begin{equation}
\delta \alpha _{2}^{0}=\partial _{\mu }k_{1}^{\mu },  \label{f24e}
\end{equation}
where $k_{1}^{\mu }$ is the invariant polynomial 
\begin{equation}
k_{1}^{\mu }=K^{\Delta }k_{\Delta }^{\mu }+K^{\Delta ^{\prime }\mu \nu
}k_{\Delta ^{\prime }\nu }+K^{\Delta ^{\prime \prime }}k_{\Delta ^{\prime
\prime }}^{\mu }.  \label{f24f}
\end{equation}

The previous results on $H\left( \delta |d\right) $ and $H^{\mathrm{inv}%
}\left( \delta |d\right) $ at strictly positive antighost numbers are
important because they control the obstructions to removing the antifields
from the first-order deformation. Indeed, due to~(\ref{f19b}--\ref{f19c})
and to the triviality of the cohomology of the exterior spacetime
differential in the space of invariant polynomials at strictly positive
antighost numbers, it follows that we can successively remove all the pieces
with $J>2$ from the non-integrated density of the first-order deformation by
adding to it only trivial terms. In conclusion we can take, without loss of
non-trivial objects, the maximum value $J=2$ of the antighost number in the
decomposition~(\ref{f18}).

\subsubsection{Determination of the first-order deformation}

For $J=2$, the first-order deformation~(\ref{f18}) reduces to 
\begin{equation}
a=a_{0}+a_{1}+a_{2},  \label{f24d}
\end{equation}
where its last representative ($\gamma a_{2}=0$) is of the form~(\ref{f19}).
The elements of pure ghost number equal to two of a basis in the space of
polynomials in $\eta ^{\alpha _{1}}$ are spanned by 
\begin{equation}
e^{2}:\left( \eta ^{a}\eta ^{b},\eta _{\mu \nu }^{a}\eta _{\rho \lambda
}^{b},C^{A}C^{B},\eta ^{a}\eta _{\mu \nu }^{b},\eta ^{a}C^{A},\eta _{\mu \nu
}^{a}C^{A}\right) ,  \label{f24g}
\end{equation}
and therefore we can write (up to $\gamma $-exact contributions) that 
\begin{eqnarray}
a_{2} &=&\frac{1}{2}\left( \alpha _{ab}\eta ^{a}\eta ^{b}+\alpha _{ab}^{\mu
\nu \rho \lambda }\eta _{\mu \nu }^{a}\eta _{\rho \lambda }^{b}+\alpha
_{AB}C^{A}C^{B}\right)  \nonumber \\
&&+\alpha _{ab}^{\mu \nu }\eta ^{a}\eta _{\mu \nu }^{b}+\alpha _{aA}\eta
^{a}C^{A}+\alpha _{aA}^{\mu \nu }\eta _{\mu \nu }^{a}C^{A}.  \label{f25}
\end{eqnarray}
According to the previous discussion, the objects $\alpha _{ab}$, $\alpha
_{ab}^{\mu \nu \rho \lambda }$, $\alpha _{AB}$, $\alpha _{ab}^{\mu \nu }$, $%
\alpha _{aA}$ and $\alpha _{aA}^{\mu \nu }$ necessarily belong to $H_{2}^{%
\mathrm{inv}}\left( \delta |d\right) $, so 
\begin{eqnarray}
\delta \alpha _{ab} &=&\partial _{\mu }k_{ab}^{\mu },\;\delta \alpha
_{ab}^{\mu \nu \rho \lambda }=\partial _{\beta }k_{ab}^{\beta \mu \nu \rho
\lambda },\;\delta \alpha _{AB}=\partial _{\mu }k_{AB}^{\mu },  \label{f27a}
\\
\delta \alpha _{ab}^{\mu \nu } &=&\partial _{\beta }k_{ab}^{\beta \mu \nu
},\;\delta \alpha _{aA}=\partial _{\mu }k_{aA}^{\mu },\;\delta \alpha
_{aA}^{\mu \nu }=\partial _{\beta }k_{aA}^{\beta \mu \nu },  \label{f27b}
\end{eqnarray}
for some currents that are invariant polynomials of antighost number one. In
addition, they are subject to the ``symmetry'' conditions (due to the
anticommutation of the ghosts) 
\begin{eqnarray}
\alpha _{ab} &=&-\alpha _{ba},\;\alpha _{ab}^{\mu \nu \rho \lambda }=-\alpha
_{ba}^{\rho \lambda \mu \nu },\;\alpha _{AB}=-\alpha _{BA},  \label{f26a} \\
\alpha _{ab}^{\mu \nu } &=&-\alpha _{ab}^{\nu \mu },\;\alpha _{aA}^{\mu \nu
}=-\alpha _{aA}^{\nu \mu },\;\alpha _{ab}^{\mu \nu \rho \lambda }=-\alpha
_{ab}^{\nu \mu \rho \lambda }=-\alpha _{ab}^{\mu \nu \lambda \rho }.
\label{f26b}
\end{eqnarray}
If we insert the expression~(\ref{f25}) into the Eq.~(\ref{f18c}) for $%
J=2$ 
\begin{equation}
\delta a_{2}+\gamma a_{1}=\partial _{\mu }j_{1}^{\mu },  \label{f27}
\end{equation}
use the formulas~(\ref{f27a}--\ref{f27b}) and recall the definitions~(\ref
{f15}--\ref{f16}), we obtain that the existence of $a_{1}$ demands further
restrictions on the currents, namely, 
\begin{eqnarray}
k_{ab}^{\beta \mu \nu \rho \lambda }\partial _{\beta }\left( \eta _{\mu \nu
}^{a}\eta _{\rho \lambda }^{b}\right) &=&\sigma _{ab}^{\nu \rho \lambda
}\left( \partial ^{\mu }\eta _{\mu \nu }^{a}\right) \eta _{\rho \lambda
}^{b}+\sigma _{ab}^{\mu \nu \lambda }\eta _{\mu \nu }^{a}\left( \partial
^{\rho }\eta _{\rho \lambda }^{b}\right) ,  \label{f28a} \\
k_{ab}^{\beta \mu \nu }\partial _{\beta }\eta _{\mu \nu }^{b} &=&\mu
_{ab}^{\nu }\partial ^{\mu }\eta _{\mu \nu }^{b},\;k_{aA}^{\beta \mu \nu
}\partial _{\beta }\eta _{\mu \nu }^{a}=\mu _{aA}^{\nu }\partial ^{\mu }\eta
_{\mu \nu }^{a},  \label{f28b}
\end{eqnarray}
for some $\sigma $ and $\mu $. On the other hand, in agreement with the
result~(\ref{f24}), every function from $H_{2}^{\mathrm{inv}}\left( \delta
|d\right) $ entering the solution~(\ref{f25}) can only be constructed out of
the three different kinds of invariant polynomials~(\ref{f21}--\ref{f23}).
Expressing now each function of the type $\alpha $ from $a_{2}$ in terms of
the allowed elements~(\ref{f21}--\ref{f23}) and imposing the supplementary
Eqs.~(\ref{f28a}--\ref{f28a}) at the level of the accompanying
currents, after some computation we infer the solutions 
\begin{eqnarray}
\alpha _{ab} &=&M_{ab}+P_{ab}=\frac{\partial W_{ab}}{\partial \varphi _{m}}%
\eta _{m}^{*}-\frac{\partial ^{2}W_{ab}}{\partial \varphi _{m}\partial
\varphi _{n}}\left( B_{m\mu \nu }\eta _{n}^{*\mu \nu }+H_{m}^{*\mu }A_{n\mu
}^{*}\right)  \nonumber \\
&&-\frac{1}{2}\frac{\partial ^{3}W_{ab}}{\partial \varphi _{m}\partial
\varphi _{n}\partial \varphi _{p}}H_{m}^{*\mu }H_{n}^{*\nu }B_{p\mu \nu
}+f_{ab}^{A}C_{A}^{*}  \nonumber \\
&&+\frac{\partial f_{ab}^{A}}{\partial \varphi _{m}}\left( V_{A}^{*\mu
}H_{m\mu }^{*}+\eta _{m}^{*\mu \nu }F_{A\mu \nu }\right) +\frac{1}{2}\frac{%
\partial ^{2}f_{ab}^{A}}{\partial \varphi _{m}\partial \varphi _{n}}%
H_{m}^{*\mu }H_{n}^{*\nu }F_{A\mu \nu },  \label{f29a}
\end{eqnarray}
\begin{equation}
\alpha _{ab}^{\mu \nu \rho \lambda }=0=\alpha _{aA}^{\mu \nu },  \label{f29b}
\end{equation}
\begin{eqnarray}
\alpha _{AB} &=&P_{AB}=f_{AB}^{C}C_{C}^{*}+\frac{\partial f_{AB}^{C}}{%
\partial \varphi _{m}}\left( V_{C}^{*\mu }H_{m\mu }^{*}+\eta _{m}^{*\mu \nu
}F_{C\mu \nu }\right)  \nonumber \\
&&+\frac{1}{2}\frac{\partial ^{2}f_{AB}^{C}}{\partial \varphi _{m}\partial
\varphi _{n}}H_{m}^{*\mu }H_{n}^{*\nu }F_{C\mu \nu },  \label{f29c}
\end{eqnarray}
\begin{equation}
\alpha _{ab}^{\mu \nu }=N_{ab}^{\mu \nu }=\frac{\partial U_{ab}}{\partial
\varphi _{m}}\eta _{m}^{*\mu \nu }+\frac{1}{2}\frac{\partial ^{2}U_{ab}}{%
\partial \varphi _{m}\partial \varphi _{n}}H_{m}^{*\mu }H_{n}^{*\nu },
\label{f29d}
\end{equation}
\begin{eqnarray}
\alpha _{aA} &=&P_{aA}=g_{aA}^{C}C_{C}^{*}+\frac{\partial g_{aA}^{C}}{%
\partial \varphi _{m}}\left( V_{C}^{*\mu }H_{m\mu }^{*}+\eta _{m}^{*\mu \nu
}F_{C\mu \nu }\right)  \nonumber \\
&&+\frac{1}{2}\frac{\partial ^{2}g_{aA}^{C}}{\partial \varphi _{m}\partial
\varphi _{n}}H_{m}^{*\mu }H_{n}^{*\nu }F_{C\mu \nu },  \label{f29e}
\end{eqnarray}
where $W_{ab}$, $f_{ab}^{A}$, $f_{AB}^{C}$, $U_{ab}$ and $g_{aA}^{C}$ depend
only on the scalar fields $\varphi _{a}$, with $W_{ab}$, $f_{ab}^{A}$ and $%
f_{AB}^{C}$ antisymmetric in their lower indices in order to enforce the
``symmetry'' properties~(\ref{f26a}--\ref{f26b}) 
\begin{equation}
W_{ab}=-W_{ba},\;f_{ab}^{A}=-f_{ba}^{A},\;f_{AB}^{C}=-f_{BA}^{C}.
\label{fas}
\end{equation}
In conclusion, the full expression of the last component from the
first-order deformation~(\ref{f24d}), such that it leads to a consistent
solution $a_{1}$ to the Eq.~(\ref{f27}), has the form~(\ref{f25}), with
the invariant polynomials $\alpha _{ab}$, $\alpha _{ab}^{\mu \nu \rho
\lambda }$, $\alpha _{AB}$, $\alpha _{ab}^{\mu \nu }$, $\alpha _{aA}$ and $%
\alpha _{aA}^{\mu \nu }$ from $H_{2}^{\mathrm{inv}}\left( \delta |d\right) $
precisely given by~(\ref{f29a}--\ref{f29e}).

With $a_{2}$ at hand, direct calculations provide the piece of antighost
number one from~(\ref{f24d}) like 
\begin{eqnarray}
a_{1} &=&\left( U_{ab}\left( B^{*a\mu \nu }\eta _{\mu \nu }^{b}+\varphi
^{*b}\eta ^{a}\right) -\frac{\partial U_{ab}}{\partial \varphi _{c}}%
H_{c}^{*\nu }\left( A^{a\mu }\eta _{\mu \nu }^{b}+H_{\nu }^{b}\eta
^{a}\right) \right)  \nonumber \\
&&+\left( f_{ab}^{A}F_{A}^{\mu \nu }-\frac{\partial W_{ab}}{\partial \varphi
_{c}}B_{c}^{\mu \nu }\right) \eta ^{a}B_{\mu \nu }^{*b}+g_{aA}^{C}F_{C}^{\mu
\nu }B_{\mu \nu }^{*a}C^{A}  \nonumber \\
&&+\left( \frac{\partial f_{ab}^{A}}{\partial \varphi _{c}}F_{A}^{\mu \nu
}H_{c\nu }^{*}-\left( \frac{\partial W_{ab}}{\partial \varphi _{c}}%
A_{c}^{*\mu }+\frac{\partial ^{2}W_{ab}}{\partial \varphi _{c}\partial
\varphi _{d}}B_{c}^{\mu \nu }H_{d\nu }^{*}+f_{ab}^{A}V_{A}^{*\mu }\right)
\right) \eta ^{a}A_{\mu }^{b}  \nonumber \\
&&+\left( -f_{AB}^{C}V_{C}^{*\mu }+\frac{\partial f_{AB}^{C}}{\partial
\varphi _{c}}F_{C}^{\mu \nu }H_{c\nu }^{*}\right) C^{A}V_{\mu }^{B} 
\nonumber \\
&&+\left( -g_{aA}^{C}V_{C}^{*\mu }+\frac{\partial g_{aA}^{C}}{\partial
\varphi _{b}}F_{C}^{\mu \nu }H_{b\nu }^{*}\right) \left( \eta ^{a}V_{\mu
}^{A}-A_{\mu }^{a}C^{A}\right) .  \label{f30}
\end{eqnarray}
The last step in completing the first-order deformation is the resolution of
the equation~(\ref{f18d}) for $k=1$ 
\begin{equation}
\delta a_{1}+\gamma a_{0}=\partial _{\mu }j_{0}^{\mu }.  \label{f30a}
\end{equation}
Evaluating $\delta a_{1}$, we find that the Eq.~(\ref{f30a}) possesses
solutions with respect to $a_{0}$ if the functions $f_{AB}^{C}$ and $%
g_{aA}^{C}$ of the undifferentiated scalar fields obey the conditions 
\begin{equation}
f_{ABC}=-f_{BAC}=-f_{ACB},\;g_{aAB}=-g_{aBA},  \label{f31}
\end{equation}
where $f_{ABC}$ and $g_{aAB}$ are defined by 
\begin{equation}
f_{ABC}=k_{AE}f_{BC}^{E},\;g_{aAB}=k_{AE}g_{aB}^{E},  \label{f32}
\end{equation}
and $k_{AE}$ denote the elements of the matrix inverse to $k^{AE}$ (used to
raise the collection indices of the vector fields). Then, we get the
interacting Lagrangian at the first order in the coupling constant like 
\begin{eqnarray}
a_{0} &=&\frac{1}{2}\left( \frac{\partial W_{ab}}{\partial \varphi _{c}}%
B_{c}^{\mu \nu }-f_{ab}^{A}F_{A}^{\mu \nu }\right) A_{\mu }^{a}A_{\nu }^{b}-%
\frac{1}{2}f_{BC}^{A}F_{A}^{\mu \nu }V_{\mu }^{B}V_{\nu }^{C}  \nonumber \\
&&-g_{aB}^{A}F_{A}^{\mu \nu }A_{\mu }^{a}V_{\nu }^{B}-U_{ab}A^{a\mu }H_{\mu
}^{b}.  \label{f33}
\end{eqnarray}

So far we have completely determined the first-order deformation of the
solution to the master equation for the model under study 
\begin{equation}
S_{1}=\int d^{2}x\left( a_{0}+a_{1}+a_{2}\right) ,  \label{f33a}
\end{equation}
where its components are expressed by~(\ref{f25}) (with the corresponding
invariant polynomials of the form~(\ref{f29a}--\ref{f29e})),~(\ref{f30}) and
(\ref{f33}). Moreover, the various functions of the undifferentiated scalar
fields are taken to obey the properties~(\ref{fas}) and~(\ref{f31}).

\subsection{Higher-order deformations}

\subsubsection{Second-order deformation}

Using the notations from~(\ref{f4}), the master equation~(\ref{mastdef})
holds to order $g^{2}$ if and only if 
\begin{equation}
\Delta =-2sb+\partial _{\mu }u^{\mu },  \label{f33b}
\end{equation}
where $\left( S_{1},S_{1}\right) =\int d^{2}x\Delta $. In other words, the
consistency of the deformed solution to the master equation at the second
order in the coupling constant requires that the integrand of $\left(
S_{1},S_{1}\right) $ should (locally) be written like a $s$-co-boundary
modulo $d$. Relying on the expression of $S_{1}$ deduced in the above, we
can emphasize a $s$-exact part in $\Delta $ if and only if the functions $%
U_{ab}$ and $W_{ab}$ coincide 
\begin{equation}
U_{ab}=W_{ab},  \label{f34}
\end{equation}
in which case we find that 
\begin{eqnarray}
\Delta  &=&-2sb+\left( t_{bcd}u^{bcd}+\frac{\partial t_{bcd}}{\partial
\varphi _{e}}u_{e}^{bcd}+\frac{\partial ^{2}t_{bcd}}{\partial \varphi
_{e}\partial \varphi _{f}}u_{ef}^{bcd}+\frac{\partial ^{3}t_{bcd}}{\partial
\varphi _{e}\partial \varphi _{f}\partial \varphi _{g}}u_{efg}^{bcd}\right) 
\nonumber \\
&&+\left( \alpha _{abc}^{A}v_{|A}^{abc}+\frac{\partial \alpha _{abc}^{A}}{%
\partial \varphi _{e}}v_{e|A}^{abc}+\frac{\partial ^{2}\alpha _{abc}^{A}}{%
\partial \varphi _{e}\partial \varphi _{f}}v_{ef|A}^{abc}\right)   \nonumber
\\
&&+\left( \alpha _{BCD}^{A}w_{|A}^{BCD}+\frac{\partial \alpha _{BCD}^{A}}{%
\partial \varphi _{e}}w_{e|A}^{BCD}+\frac{\partial ^{2}\alpha _{BCD}^{A}}{%
\partial \varphi _{e}\partial \varphi _{f}}w_{ef|A}^{BCD}\right)   \nonumber
\\
&&+\left( \alpha _{abB}^{A}z_{|A}^{abB}+\frac{\partial \alpha _{abB}^{A}}{%
\partial \varphi _{e}}z_{e|A}^{abB}+\frac{\partial ^{2}\alpha _{abB}^{A}}{%
\partial \varphi _{e}\partial \varphi _{f}}z_{ef|A}^{abB}\right)   \nonumber
\\
&&+\left( \alpha _{aBC}^{A}q_{|A}^{aBC}+\frac{\partial \alpha _{aBC}^{A}}{%
\partial \varphi _{e}}q_{e|A}^{aBC}+\frac{\partial ^{2}\alpha _{aBC}^{A}}{%
\partial \varphi _{e}\partial \varphi _{f}}q_{ef|A}^{aBC}\right) .
\label{f36}
\end{eqnarray}
The expression of $b$ in~(\ref{f36}) is 
\begin{equation}
b=-\frac{1}{4}Q_{\mu \nu }^{A}k_{AB}Q^{B\mu \nu },  \label{f42}
\end{equation}
where we performed the notation 
\begin{eqnarray}
Q_{\mu \nu }^{A} &=&\left( \frac{\partial f_{ab}^{A}}{\partial \varphi _{c}}%
\eta _{c\mu \nu }^{*}+\frac{1}{2}\frac{\partial ^{2}f_{ab}^{A}}{\partial
\varphi _{c}\partial \varphi _{d}}H_{c\mu }^{*}H_{d\nu }^{*}\right) \eta
^{a}\eta ^{b}-\left( f_{ab}^{A}A_{\mu }^{a}A_{\nu }^{b}+f_{BC}^{A}V_{\mu
}^{B}V_{\nu }^{C}\right)   \nonumber \\
&&-\left( \frac{\partial f_{ab}^{A}}{\partial \varphi _{c}}H_{c[\mu
}^{*}A_{\nu ]}^{b}+2f_{ab}^{A}B_{\mu \nu }^{*b}\right) \eta ^{a}-\frac{%
\partial f_{BC}^{A}}{\partial \varphi _{c}}H_{c[\mu }^{*}V_{\nu
]}^{C}C^{B}-g_{aB}^{A}A_{[\mu }^{a}V_{\nu ]}^{B}  \nonumber \\
&&+\left( \frac{\partial f_{BC}^{A}}{\partial \varphi _{c}}\eta _{c\mu \nu
}^{*}+\frac{1}{2}\frac{\partial ^{2}f_{BC}^{A}}{\partial \varphi
_{c}\partial \varphi _{d}}H_{c\mu }^{*}H_{d\nu }^{*}\right)
C^{B}C^{C}+2g_{aB}^{A}B_{\mu \nu }^{*a}C^{B}  \nonumber \\
&&+2\left( \frac{\partial g_{aB}^{A}}{\partial \varphi _{c}}\eta _{c\mu \nu
}^{*}+\frac{1}{2}\frac{\partial ^{2}g_{aB}^{A}}{\partial \varphi
_{c}\partial \varphi _{d}}H_{c\mu }^{*}H_{d\nu }^{*}\right) \eta ^{a}C^{B} 
\nonumber \\
&&+\frac{\partial g_{aB}^{A}}{\partial \varphi _{c}}\left( H_{c[\mu
}^{*}A_{\nu ]}^{a}C^{B}-H_{c[\mu }^{*}V_{\nu ]}^{B}\eta ^{a}\right) .
\label{f43}
\end{eqnarray}
The coefficients denoted by $t$ and $\alpha $ from~(\ref{f36}) involve only
the undifferentiated scalar fields and are given by 
\begin{equation}
t_{abc}=W_{e[a}\frac{\partial W_{bc]}}{\partial \varphi _{e}},  \label{f35a}
\end{equation}
\begin{equation}
\alpha _{abc}^{A}=f_{e[a}^{A}\frac{\partial W_{bc]}}{\partial \varphi _{e}}%
+W_{e[a}\frac{\partial f_{bc]}^{A}}{\partial \varphi _{e}}%
-f_{[ab}^{E}g_{c]E}^{A},  \label{f35b}
\end{equation}
\begin{equation}
\alpha _{BCD}^{A}=f_{E[B}^{A}f_{CD]}^{E},  \label{f35c}
\end{equation}
\begin{equation}
\alpha _{abB}^{A}=g_{eB}^{A}\frac{\partial W_{ab}}{\partial \varphi _{e}}%
+f_{EB}^{A}f_{ab}^{E}+W_{e[a}\frac{\partial g_{b]B}^{A}}{\partial \varphi
_{e}}+\left( g_{aB}^{E}g_{bE}^{A}-g_{bB}^{E}g_{aE}^{A}\right) ,  \label{f35d}
\end{equation}
\begin{equation}
\alpha _{aBC}^{A}=g_{aE}^{A}f_{BC}^{E}-W_{ea}\frac{\partial f_{BC}^{A}}{%
\partial \varphi _{e}}+\left(
f_{EB}^{A}g_{aC}^{E}-f_{EC}^{A}g_{aB}^{E}\right) ,  \label{f35e}
\end{equation}
while the functions of the type $u$, $v$, $w$, $z$ and $q$ contain only
undifferentiated ghosts and antifields. Their expressions are listed below 
\begin{equation}
u^{bcd}=\left( A^{b\mu }A^{c\nu }-B^{*b\mu \nu }\eta ^{c}\right) \eta _{\mu
\nu }^{d}-\left( A^{b\mu }H_{\mu }^{d}+\varphi ^{*b}\eta ^{d}\right) \eta
^{c},  \label{f37a}
\end{equation}
\begin{eqnarray}
u_{e}^{bcd} &=&\left( B_{e}^{\mu \nu }A_{\mu }^{b}A_{\nu }^{d}+A_{e}^{*\mu
}A_{\mu }^{b}\eta ^{d}\right) \eta ^{c}+\left( B_{e}^{\mu \nu }B_{\mu \nu
}^{*b}-H_{e}^{*\mu }H_{\mu }^{b}-\frac{1}{3}\eta _{e}^{*}\eta ^{b}\right)
\eta ^{c}\eta ^{d}  \nonumber \\
&&+\left( H_{e}^{*\nu }A^{b\mu }-\eta _{e}^{*\mu \nu }\eta ^{b}\right) \eta
^{c}\eta _{\mu \nu }^{d},  \label{f37b}
\end{eqnarray}
\begin{equation}
u_{ef}^{bcd}=\left( \eta _{e}^{*\mu \nu }B_{f\mu \nu }\eta ^{b}-B_{e\mu \nu
}H_{f}^{*\nu }A^{b\mu }-\frac{1}{2}H_{e}^{*\mu }H_{f}^{*\nu }\eta _{\mu \nu
}^{b}-A_{e\mu }^{*}H_{f}^{*\mu }\eta ^{b}\right) \eta ^{c}\eta ^{d},
\label{f37c}
\end{equation}
\begin{equation}
u_{efg}^{bcd}=\frac{1}{6}B_{e\mu \nu }H_{f}^{*\mu }H_{g}^{*\nu }\eta
^{b}\eta ^{c}\eta ^{d},  \label{f37d}
\end{equation}
\begin{equation}
v_{|A}^{abc}=-\left( \frac{1}{3}C_{A}^{*}\eta ^{a}+V_{A}^{*\mu }A_{\mu
}^{a}+F_{A}^{\mu \nu }B_{\mu \nu }^{*a}\right) \eta ^{b}\eta ^{c}+F_{A}^{\mu
\nu }\eta ^{a}A_{\mu }^{b}A_{\nu }^{c},  \label{f38a}
\end{equation}
\begin{equation}
v_{e|A}^{abc}=-\left( \frac{1}{3}V_{A}^{*\mu }H_{e\mu }^{*}\eta ^{a}+\frac{1%
}{3}F_{A}^{\mu \nu }\eta _{e\mu \nu }^{*}\eta ^{a}+F_{A}^{\mu \nu }H_{e\mu
}^{*}A_{\nu }^{a}\right) \eta ^{b}\eta ^{c},  \label{f38b}
\end{equation}
\begin{equation}
v_{ef|A}^{abc}=-\frac{1}{6}F_{A}^{\mu \nu }H_{e\mu }^{*}H_{f\nu }^{*}\eta
^{a}\eta ^{b}\eta ^{c},  \label{f38c}
\end{equation}
\begin{equation}
w_{|A}^{BCD}=-\left( \frac{1}{3}C_{A}^{*}C^{B}+V_{A}^{*\mu }V_{\mu
}^{B}\right) C^{C}C^{D}+F_{A}^{\mu \nu }C^{B}V_{\mu }^{C}V_{\nu }^{D},
\label{f39a}
\end{equation}
\begin{equation}
w_{e|A}^{BCD}=-\left( \frac{1}{3}V_{A}^{*\mu }H_{e\mu }^{*}C^{B}+\frac{1}{3}%
F_{A}^{\mu \nu }\eta _{e\mu \nu }^{*}C^{B}+F_{A}^{\mu \nu }H_{e\mu
}^{*}V_{\nu }^{B}\right) C^{C}C^{D},  \label{f39b}
\end{equation}
\begin{equation}
w_{ef|A}^{BCD}=-\frac{1}{6}F_{A}^{\mu \nu }H_{e\mu }^{*}H_{f\nu
}^{*}C^{B}C^{C}C^{D},  \label{f39c}
\end{equation}
\begin{eqnarray}
z_{|A}^{abB} &=&\left( F_{A}^{\mu \nu }A_{\mu }^{a}A_{\nu }^{b}+2V_{A}^{*\mu
}A_{\mu }^{b}\eta ^{a}-2F_{A}^{\mu \nu }B_{\mu \nu }^{*a}\eta ^{b}\right)
C^{B}  \nonumber \\
&&-F_{A}^{\mu \nu }V_{[\mu }^{B}A_{\nu ]}^{b}\eta ^{a}-\left(
C_{A}^{*}C^{B}+V_{A}^{*\mu }V_{\mu }^{B}\right) \eta ^{a}\eta ^{b},
\label{f40a}
\end{eqnarray}
\begin{eqnarray}
z_{e|A}^{abB} &=&-\left( V_{A}^{*\mu }H_{e\mu }^{*}C^{B}+F_{A}^{\mu \nu
}\eta _{e\mu \nu }^{*}C^{B}+F_{A}^{\mu \nu }H_{e\mu }^{*}V_{\nu }^{B}\right)
\eta ^{a}\eta ^{b}  \nonumber \\
&&-F_{A}^{\mu \nu }H_{e[\mu }^{*}A_{\nu ]}^{b}C^{B}\eta ^{a},  \label{f40b}
\end{eqnarray}
\begin{equation}
z_{ef|A}^{abB}=-\frac{1}{2}F_{A}^{\mu \nu }H_{e\mu }^{*}H_{f\nu
}^{*}C^{B}\eta ^{a}\eta ^{b},  \label{f40c}
\end{equation}
\begin{eqnarray}
q_{|A}^{aBC} &=&\left( C_{A}^{*}\eta ^{a}+V_{A}^{*\mu }A_{\mu
}^{a}+F_{A}^{\mu \nu }B_{\mu \nu }^{*a}\right) C^{B}C^{C}  \nonumber \\
&&-\left( 2V_{A}^{*\mu }V_{\mu }^{C}C^{B}+F_{A}^{\mu \nu }V_{\mu }^{B}V_{\nu
}^{C}\right) \eta ^{a}+F_{A}^{\mu \nu }A_{[\mu }^{a}V_{\nu ]}^{C}C^{B},
\label{f41a}
\end{eqnarray}
\begin{eqnarray}
q_{e|A}^{aBC} &=&\left( F_{A}^{\mu \nu }\eta _{e\mu \nu }^{*}\eta
^{a}+F_{A}^{\mu \nu }H_{e\mu }^{*}A_{\nu }^{a}+V_{A}^{*\mu }H_{e\mu
}^{*}\eta ^{a}\right) C^{B}C^{C}  \nonumber \\
&&+F_{A}^{\mu \nu }H_{e[\mu }^{*}V_{\nu ]}^{C}\eta ^{a}C^{B},  \label{f41b}
\end{eqnarray}
\begin{equation}
q_{ef|A}^{aBC}=\frac{1}{2}F_{A}^{\mu \nu }H_{e\mu }^{*}H_{f\nu }^{*}\eta
^{a}C^{B}C^{C}.  \label{f41c}
\end{equation}
Since none of the terms from the right-hand side of~(\ref{f36}) containing
the quantities $t_{abc}$, $\alpha _{abc}^{A}$, $\alpha _{BCD}^{A}$, $\alpha
_{abB}^{A}$, $\alpha _{aBC}^{A}$ and their derivatives with respect to the
scalar fields can be written in a $s$-exact modulo $d$ form, it follows that
the second-order deformation of the solution to the master equation exists
if and only if all these vanish 
\begin{equation}
t_{abc}=0,\;\alpha _{abc}^{A}=0,\;\alpha _{BCD}^{A}=0,\;\alpha
_{abB}^{A}=0,\;\alpha _{aBC}^{A}=0.  \label{relcoef}
\end{equation}
In conclusion, the deformed solution to the master equation is consistent to
order $g^{2}$ if and only if the functions $W_{ab}$, $f_{BC}^{A}$, $%
f_{ab}^{A}$ and $g_{aB}^{A}$ satisfy the Eqs.~(\ref{relcoef}), in which
case the second-order deformation is expressed like in~(\ref{f42}). We will
comment more on the Eqs.~(\ref{relcoef}) in Section 5, where we
explicitly compute some particular solutions, which allow a nice geometric
and algebraic interpretation.

\subsubsection{Third- and higher-order deformations}

If we denote the third-order deformation by $S_{3}=\int d^{2}x\,c$, the
master equation~(\ref{mastdef}) holds to order $g^{3}$ if and only if 
\begin{equation}
\Lambda =-sc+\partial _{\mu }w^{\mu },  \label{f44}
\end{equation}
where$\left( S_{1},S_{2}\right) =\int d^{2}x\,\Lambda $. Taking into account
the expressions of the first- and second-order deformations obtained in the
above, after some computation we infer that 
\begin{eqnarray}
\Lambda &=&\left( \eta ^{a}g_{aE}^{A}+\frac{1}{2}C^{D}f_{DE}^{A}\right)
Q^{E\mu \nu }k_{AB}Q_{\mu \nu }^{B}  \nonumber \\
&&+\left( \alpha _{abc}^{A}\bar{v}_{|A}^{abc}+\frac{\partial \alpha
_{abc}^{A}}{\partial \varphi _{e}}\bar{v}_{e|A}^{abc}+\frac{\partial
^{2}\alpha _{abc}^{A}}{\partial \varphi _{e}\partial \varphi _{f}}\bar{v}%
_{ef|A}^{abc}\right)  \nonumber \\
&&+\left( \alpha _{BCD}^{A}\bar{w}_{|A}^{BCD}+\frac{\partial \alpha
_{BCD}^{A}}{\partial \varphi _{e}}\bar{w}_{e|A}^{BCD}+\frac{\partial
^{2}\alpha _{BCD}^{A}}{\partial \varphi _{e}\partial \varphi _{f}}\bar{w}%
_{ef|A}^{BCD}\right)  \nonumber \\
&&+\left( \alpha _{abB}^{A}\bar{z}_{|A}^{abB}+\frac{\partial \alpha
_{abB}^{A}}{\partial \varphi _{e}}\bar{z}_{e|A}^{abB}+\frac{\partial
^{2}\alpha _{abB}^{A}}{\partial \varphi _{e}\partial \varphi _{f}}\bar{z}%
_{ef|A}^{abB}\right)  \nonumber \\
&&+\left( \alpha _{aBC}^{A}\bar{q}_{|A}^{aBC}+\frac{\partial \alpha
_{aBC}^{A}}{\partial \varphi _{e}}\bar{q}_{e|A}^{aBC}+\frac{\partial
^{2}\alpha _{aBC}^{A}}{\partial \varphi _{e}\partial \varphi _{f}}\bar{q}%
_{ef|A}^{aBC}\right) ,  \label{f46}
\end{eqnarray}
where we employed the notations 
\begin{equation}
\bar{v}_{|A}^{abc}=\frac{1}{2}k_{AM}Q^{M\mu \nu }\left( B_{\mu \nu
}^{*a}\eta ^{a}\eta ^{b}-\eta ^{a}A_{\mu }^{b}A_{\nu }^{c}\right) ,
\label{f47a}
\end{equation}
\begin{equation}
\bar{v}_{e|A}^{abc}=\frac{1}{2}k_{AM}Q^{M\mu \nu }\left( \frac{1}{3}\eta
_{e\mu \nu }^{*}\eta ^{a}+\frac{1}{2}H_{e[\mu }^{*}A_{\nu ]}^{a}\right) \eta
^{b}\eta ^{c},  \label{f47b}
\end{equation}
\begin{equation}
\bar{v}_{ef|A}^{abc}=\frac{1}{12}k_{AM}Q^{M\mu \nu }H_{e\mu }^{*}H_{f\nu
}^{*}\eta ^{a}\eta ^{b}\eta ^{c},  \label{f47c}
\end{equation}
\begin{equation}
\bar{w}_{|A}^{BCD}=-\frac{1}{2}k_{AM}Q^{M\mu \nu }C^{B}V_{\mu }^{C}V_{\nu
}^{D},  \label{f48a}
\end{equation}
\begin{equation}
\bar{w}_{e|A}^{BCD}=\frac{1}{2}k_{AM}Q^{M\mu \nu }\left( \frac{1}{3}\eta
_{e\mu \nu }^{*}C^{B}+\frac{1}{2}H_{e[\mu }^{*}V_{\nu ]}^{B}\right)
C^{C}C^{D},  \label{f48b}
\end{equation}
\begin{equation}
\bar{w}_{ef|A}^{BCD}=\frac{1}{12}k_{AM}Q^{M\mu \nu }H_{e\mu }^{*}H_{f\nu
}^{*}C^{B}C^{C}C^{D},  \label{f48c}
\end{equation}
\begin{equation}
\bar{z}_{|A}^{abB}=-\frac{1}{2}k_{AM}Q^{M\mu \nu }\left( \eta ^{a}A_{[\mu
}^{b}V_{\nu ]}^{B}+A_{\mu }^{a}A_{\nu }^{b}C^{B}+2B_{\mu \nu }^{*b}\eta
^{a}C^{B}\right) ,  \label{f49a}
\end{equation}
\begin{equation}
\bar{z}_{e|A}^{abB}=\frac{1}{2}k_{AM}Q^{M\mu \nu }\left[ \left( \eta _{e\mu
\nu }^{*}C^{B}+\frac{1}{2}H_{e[\mu }^{*}V_{\nu ]}^{B}\right) \eta ^{a}\eta
^{b}-\frac{1}{2}H_{e[\mu }^{*}A_{\nu ]}^{b}\eta ^{a}C^{B}\right] ,
\label{f49b}
\end{equation}
\begin{equation}
\bar{z}_{ef|A}^{abB}=\frac{1}{4}k_{AM}Q^{M\mu \nu }k_{AM}Q^{M\mu \nu
}H_{e\mu }^{*}H_{f\nu }^{*}\eta ^{a}\eta ^{b}C^{B},  \label{f49c}
\end{equation}
\begin{equation}
\bar{q}_{|A}^{aBC}=-\frac{1}{2}k_{AM}Q^{M\mu \nu }\left( A_{[\mu }^{a}V_{\nu
]}^{C}C^{B}+B_{\mu \nu }^{*a}C^{B}C^{C}-\eta ^{a}V_{\mu }^{B}V_{\nu
}^{C}\right) ,  \label{f50a}
\end{equation}
\begin{equation}
\bar{q}_{e|A}^{aBC}=-\frac{1}{2}k_{AM}Q^{M\mu \nu }\left[ \left( \eta _{e\mu
\nu }^{*}\eta ^{a}+H_{e[\mu }^{*}A_{\nu ]}^{a}\right) C^{B}C^{C}+H_{e[\mu
}^{*}V_{\nu ]}^{C}\eta ^{a}C^{B}\right] ,  \label{f50b}
\end{equation}
\begin{equation}
\bar{q}_{ef|A}^{aBC}=-\frac{1}{4}k_{AM}Q^{M\mu \nu }k_{AM}Q^{M\mu \nu
}H_{e\mu }^{*}H_{f\nu }^{*}\eta ^{a}C^{B}C^{C}.  \label{f50c}
\end{equation}
It is now clear that none of the terms in the right-hand side of~(\ref{f46})
can be written like in~(\ref{f44}). However, if we take into account the
Eqs.~(\ref{relcoef}) and the antisymmetry properties~(\ref{f31}) of the
functions~(\ref{f32}), we find that $\Lambda =0$, so we can take $c=0$ in (%
\ref{f44}), and consequently obtain that 
\begin{equation}
S_{3}=0.  \label{f100}
\end{equation}
The equation that governs the fourth-order deformation $S_{4}$ reads as 
\begin{equation}
sS_{4}+\left( S_{3},S_{1}\right) +\frac{1}{2}\left( S_{2},S_{2}\right) =0.
\label{f100a}
\end{equation}
On the one hand, the result~(\ref{f100}) implies that $\left(
S_{3},S_{1}\right) =0$ and, on the other hand, if we compute $\left(
S_{2},S_{2}\right) $, where $S_{2}=\int d^{2}x\,b$, with $b$ given in~(\ref
{f42}), it follows that $\left( S_{2},S_{2}\right) =0$, so we can set 
\begin{equation}
S_{4}=0.  \label{f100b}
\end{equation}
Meanwhile, we remark that the equations responsible for the deformations $%
\left( S_{k}\right) _{k>4}$ involve only the solutions $\left( S_{j}\right)
_{j\geq 3}$, which further allows us to put 
\begin{equation}
S_{k}=0,\;k>4.  \label{f100c}
\end{equation}
In conclusion, among the higher-order deformations of the solution to the
master equation, only that of second-order is non-vanishing and non-trivial.

\section{Identification of the interacting theory}

Putting together the results deduced in the previous section, we can write
the full deformed solution to the master equation~(\ref{mastdef}), that is
consistent to all orders in the coupling constant, under the form 
\begin{eqnarray}
\bar{S} &=&S+gS_{1}+g^{2}S_{2}=\int d^{2}x\left( H_{\mu }^{a}D^{\mu }\varphi
_{a}+\frac{1}{2}B_{a}^{\mu \nu }\bar{F}_{\mu \nu }^{^{\prime }a}+A_{a}^{*\mu
}\left( D_{\mu }\right) _{\;\;b}^{a}\eta ^{b}\right.  \nonumber \\
&&-gW_{ab}\varphi ^{*a}\eta ^{b}+gB^{*a\mu \nu }\left( W_{ab}\eta _{\mu \nu
}^{b}-\frac{\partial W_{ab}}{\partial \varphi _{c}}B_{c\mu \nu }\eta
^{b}\right)  \nonumber \\
&&+H_{a}^{*\mu }\left( \left( D^{\nu }\right) _{\;\;b}^{a}\eta _{\mu \nu
}^{b}-g\left( \frac{\partial W_{bc}}{\partial \varphi _{a}}H_{\mu }^{c}-%
\frac{\partial ^{2}W_{bc}}{\partial \varphi _{a}\partial \varphi _{d}}%
B_{d\mu \nu }A^{c\nu }\right) \eta ^{b}\right)  \nonumber \\
&&+V_{A}^{*\mu }\left( \left( D_{\mu }\right) _{\;\;B}^{A}C^{B}+\left(
D_{\mu }\right) _{\;\;a}^{A}\eta ^{a}\right)  \nonumber \\
&&-\frac{1}{4}\left( F_{\mu \nu }^{A}-Q_{\mu \nu }^{A}\right) k_{AB}\left(
F^{B\mu \nu }-Q^{B\mu \nu }\right)  \nonumber \\
&&+\frac{g}{2}\left( f_{ab}^{A}C_{A}^{*}+\frac{\partial W_{ab}}{\partial
\varphi _{c}}\eta _{c}^{*}+\frac{\partial f_{ab}^{A}}{\partial \varphi _{c}}%
V_{A}^{*\mu }H_{c\mu }^{*}+\frac{\partial ^{2}W_{ab}}{\partial \varphi
_{c}\partial \varphi _{d}}A_{d}^{*\mu }H_{c\mu }^{*}\right.  \nonumber \\
&&\left. -\frac{\partial ^{2}W_{ab}}{\partial \varphi _{c}\partial \varphi
_{d}}B_{d\mu \nu }\eta _{c}^{*\mu \nu }-\frac{1}{2}\frac{\partial ^{3}W_{ab}%
}{\partial \varphi _{c}\partial \varphi _{d}\partial \varphi _{e}}B_{c\mu
\nu }H_{d}^{*\mu }H_{e}^{*\nu }\right) \eta ^{a}\eta ^{b}  \nonumber \\
&&+g\left( \frac{\partial W_{ab}}{\partial \varphi _{c}}\eta _{c}^{*\mu \nu
}+\frac{1}{2}\frac{\partial ^{2}W_{ab}}{\partial \varphi _{c}\partial
\varphi _{d}}H_{c}^{*\mu }H_{d}^{*\nu }\right) \eta ^{a}\eta _{\mu \nu }^{b}
\nonumber \\
&&+\frac{g}{2}\left( f_{BC}^{A}C_{A}^{*}+\frac{\partial f_{BC}^{A}}{\partial
\varphi _{c}}V_{A}^{*\mu }H_{c\mu }^{*}\right) C^{B}C^{C}  \nonumber \\
&&\left. +g\left( g_{aB}^{A}C_{A}^{*}+\frac{\partial g_{aB}^{A}}{\partial
\varphi _{b}}V_{A}^{*\mu }H_{b\mu }^{*}\right) \eta ^{a}C^{B}\right) ,
\label{f51}
\end{eqnarray}
where we performed the notations 
\begin{eqnarray}
D^{\mu }\varphi _{a} &=&\partial ^{\mu }\varphi _{a}+gW_{ab}A^{b\mu },
\label{f52a} \\
\bar{F}_{\mu \nu }^{^{\prime }a} &=&\partial _{[\mu }A_{\nu ]}^{a}+g\frac{%
\partial W_{bc}}{\partial \varphi _{a}}A_{\mu }^{b}A_{\nu }^{c},
\label{f52e} \\
\left( D^{\mu }\right) _{\;\;b}^{a} &=&\delta _{\;\;b}^{a}\partial ^{\mu }-g%
\frac{\partial W_{bc}}{\partial \varphi _{a}}A_{\mu }^{c},  \label{f52b} \\
\left( D_{\mu }\right) _{\;\;B}^{A} &=&\delta _{\;\;B}^{A}\partial _{\mu
}-g\left( f_{BC}^{A}V_{\mu }^{C}-g_{aB}^{A}A_{\mu }^{a}\right) ,
\label{f52c} \\
\left( D_{\mu }\right) _{\;\;a}^{A} &=&-g\left( f_{ab}^{A}A_{\mu
}^{b}+g_{aB}^{A}V_{\mu }^{B}\right) .  \label{f52d}
\end{eqnarray}
We note that the deformed solution~(\ref{f51}) contains components of
antighost numbers ranging from zero to four, unlike the solution~(\ref{f5})
of the master equation for the free model, which stopped at antighost number
one. We stress again that the coefficients $W_{ab}$, $f_{ab}^{A}$, $%
f_{BC}^{A}$ and $g_{aB}^{A}$ are all functions of the undifferentiated
scalar fields, that \textit{must} obey the antisymmetry properties~(\ref{fas}%
) and~(\ref{f31}), as well as the Eqs.~(\ref{relcoef}), where the
functions $t_{abc}$, $\alpha _{abc}^{A}$, $\alpha _{BCD}^{A}$, $\alpha
_{abB}^{A}$ and $\alpha _{aBC}^{A}$ are defined in the formulas~(\ref{f35a}--%
\ref{f35e}).

At this stage, we have all the information necessary at the identification
of the interacting gauge theory behind our deformation procedure. According
to the general rules of the antifield-BRST formalism, the Lagrangian action
that describes the coupled model is nothing but the antighost number zero
piece from~(\ref{f51}), so it has the expression 
\begin{equation}
\bar{S}_{0}\left[ A_{\mu }^{a},H_{\mu }^{a},\varphi _{a},B_{a}^{\mu \nu
},V_{\mu }^{A}\right] =\int d^{2}x\left( H_{\mu }^{a}D^{\mu }\varphi _{a}+%
\frac{1}{2}B_{a}^{\mu \nu }\bar{F}_{\mu \nu }^{\prime a}-\frac{1}{4}\bar{F}%
_{\mu \nu }^{A}\bar{F}_{A}^{\mu \nu }\right) ,  \label{f53}
\end{equation}
where we used the notation 
\begin{equation}
\bar{F}_{\mu \nu }^{A}=\partial _{[\mu }V_{\nu ]}^{A}+g\left(
f_{BC}^{A}V_{\mu }^{B}V_{\nu }^{C}+f_{ab}^{A}A_{\mu }^{a}A_{\nu
}^{b}+g_{aB}^{A}A_{[\mu }^{a}V_{\nu ]}^{B}\right) .  \label{f54}
\end{equation}
The terms of antighost number one from~(\ref{f51}) offer us the generating
set of deformed gauge transformations corresponding to the Lagrangian action
(\ref{f53}), or, in other words, the gauge symmetries of the interacting
action, namely, 
\begin{equation}
\bar{\delta}_{\epsilon }A_{\mu }^{a}=\left( D_{\mu }\right)
_{\;\;b}^{a}\epsilon ^{b},  \label{f55a}
\end{equation}
\begin{eqnarray}
\bar{\delta}_{\epsilon }H_{\mu }^{a} &=&\left( D^{\nu }\right)
_{\;\;b}^{a}\epsilon _{\mu \nu }^{b}-g\left( \frac{\partial W_{bc}}{\partial
\varphi _{a}}H_{\mu }^{c}-\frac{\partial ^{2}W_{bc}}{\partial \varphi
_{a}\partial \varphi _{d}}B_{d\mu \nu }A^{c\nu }\right) \epsilon ^{b} 
\nonumber \\
&&+\bar{F}_{A\mu \nu }\left( \frac{\partial \left( D^{\nu }\right)
_{\;\;b}^{A}}{\partial \varphi _{a}}\epsilon ^{b}+\frac{\partial \left(
D^{\nu }\right) _{\;\;B}^{A}}{\partial \varphi _{a}}\epsilon ^{B}\right) ,
\label{f55b}
\end{eqnarray}
\begin{equation}
\bar{\delta}_{\epsilon }\varphi _{a}=-gW_{ab}\epsilon ^{b},\;\bar{\delta}%
_{\epsilon }V_{\mu }^{A}=\left( D_{\mu }\right) _{\;\;B}^{A}\epsilon
^{B}+\left( D_{\mu }\right) _{\;\;a}^{A}\epsilon ^{a},  \label{f55c}
\end{equation}
\begin{equation}
\bar{\delta}_{\epsilon }B_{a}^{\mu \nu }=g\left( W_{ab}\epsilon ^{b\mu \nu }-%
\frac{\partial W_{ab}}{\partial \varphi _{c}}B_{c}^{\mu \nu }\epsilon
^{b}\right) +g\left( g_{aB}^{A}\epsilon ^{B}+f_{ab}^{A}\epsilon ^{b}\right) 
\bar{F}_{A}^{\mu \nu }.  \label{f55d}
\end{equation}

There also appear two types of antighost number two elements in~(\ref{f51}).
As we have stated in the end of Section 2, their presence indicates that the
gauge algebra associated with the deformed gauge transformations is open, so
the commutators among the gauge transformations~(\ref{f55a}--\ref{f55d})
only close on the stationary surface of the field equations corresponding to
the action~(\ref{f53}). Indeed, let $\epsilon ^{\alpha _{1}}=\left( \epsilon
^{a},\epsilon _{\mu \nu }^{a},\epsilon ^{A}\right) $ and $\xi ^{\alpha
_{1}}=\left( \xi ^{a},\xi _{\mu \nu }^{a},\xi ^{A}\right) $ be two different
sets of gauge parameters. Then, the expressions of the commutators between
the deformed gauge transformations~(\ref{f55a}--\ref{f55d}) associated with
these parameters are completely determined from the antighost number two
objects in~(\ref{f51}) under the form 
\begin{equation}
\left[ \bar{\delta}_{\epsilon },\bar{\delta}_{\xi }\right] \varphi _{a}=\bar{%
\delta}_{\Lambda }\varphi _{a},  \label{com1}
\end{equation}
\begin{equation}
\left[ \bar{\delta}_{\epsilon },\bar{\delta}_{\xi }\right] A_{\mu }^{a}=\bar{%
\delta}_{\Lambda }A_{\mu }^{a}+g\frac{\delta \bar{S}_{0}}{\delta H^{d\mu }}%
\frac{\partial ^{2}W_{bc}}{\partial \varphi _{a}\partial \varphi _{d}}%
\epsilon ^{b}\xi ^{c},  \label{com2}
\end{equation}
\begin{eqnarray}
&&\left[ \bar{\delta}_{\epsilon },\bar{\delta}_{\xi }\right] B_{a}^{\mu \nu
}=\bar{\delta}_{\Lambda }B_{a}^{\mu \nu }+2g^{2}\frac{\delta \bar{S}_{0}}{%
\delta B_{d\mu \nu }}k_{AB}\left( \left(
g_{aC}^{A}g_{dD}^{B}-g_{aD}^{A}g_{dC}^{B}\right) \epsilon ^{C}\xi ^{D}\right.
\nonumber \\
&&\left. +f_{a[b}^{A}f_{c]d}^{B}\epsilon ^{b}\xi ^{c}-\left(
g_{aC}^{A}f_{bd}^{B}+f_{ab}^{A}g_{dC}^{B}\right) \left( \epsilon ^{b}\xi
^{C}-\xi ^{b}\epsilon ^{C}\right) \right)  \nonumber \\
&&+g^{2}\delta _{\alpha }^{[\mu }\delta _{\beta }^{\nu ]}\frac{\delta \bar{S}%
_{0}}{\delta H_{\beta }^{d}}k_{AB}\left( A^{b\alpha }\left( g_{aC}^{A}\frac{%
\partial g_{bD}^{B}}{\partial \varphi _{d}}-g_{aD}^{A}\frac{\partial
g_{bC}^{B}}{\partial \varphi _{d}}\right) \epsilon ^{C}\xi ^{D}\right. 
\nonumber \\
&&-V^{E\alpha }g_{a[C}^{A}\frac{\partial f_{D]E}^{B}}{\partial \varphi _{d}}%
\epsilon ^{C}\xi ^{D}-\left( A^{e\alpha }f_{a[b}^{A}\frac{\partial
f_{c]e}^{B}}{\partial \varphi _{d}}+V^{E\alpha }f_{a[b}^{A}\frac{\partial
g_{c]E}^{B}}{\partial \varphi _{d}}\right) \epsilon ^{b}\xi ^{c}  \nonumber
\\
&&+\left( A^{c\alpha }\left( g_{aC}^{A}\frac{\partial f_{bc}^{B}}{\partial
\varphi _{d}}+f_{ab}^{A}\frac{\partial g_{cC}^{B}}{\partial \varphi _{d}}%
\right) \right.  \nonumber \\
&&\left. \left. -V^{D\alpha }\left( g_{aC}^{A}\frac{\partial g_{bD}^{B}}{%
\partial \varphi _{d}}-f_{ab}^{A}\frac{\partial f_{CD}^{B}}{\partial \varphi
_{d}}\right) \right) \left( \epsilon ^{b}\xi ^{C}-\xi ^{b}\epsilon
^{C}\right) \right) ,  \label{com3}
\end{eqnarray}
\begin{eqnarray}
&&\left[ \bar{\delta}_{\epsilon },\bar{\delta}_{\xi }\right] H_{\mu }^{a}=%
\bar{\delta}_{\Lambda }H_{\mu }^{a}-g\frac{\delta \bar{S}_{0}}{\delta
A^{d\mu }}\frac{\partial ^{2}W_{bc}}{\partial \varphi _{a}\partial \varphi
_{d}}\epsilon ^{b}\xi ^{c}  \nonumber \\
&&+g\frac{\delta \bar{S}_{0}}{\delta V^{A\mu }}\left( \frac{\partial
f_{BC}^{A}}{\partial \varphi _{a}}\epsilon ^{B}\xi ^{C}+\frac{\partial
f_{bc}^{A}}{\partial \varphi _{a}}\epsilon ^{b}\xi ^{c}+\frac{\partial
g_{bB}^{A}}{\partial \varphi _{a}}\left( \epsilon ^{b}\xi ^{B}-\xi
^{b}\epsilon ^{B}\right) \right)  \nonumber \\
&&+g^{2}\frac{\delta \bar{S}_{0}}{\delta B_{d}^{\mu \alpha }}k_{AB}\left(
A^{b\alpha }\left( g_{dC}^{A}\frac{\partial g_{bD}^{B}}{\partial \varphi _{a}%
}-g_{dD}^{A}\frac{\partial g_{bC}^{B}}{\partial \varphi _{a}}\right)
\epsilon ^{C}\xi ^{D}\right.  \nonumber \\
&&-V^{E\alpha }g_{d[C}^{A}\frac{\partial f_{D]E}^{B}}{\partial \varphi _{a}}%
\epsilon ^{C}\xi ^{D}-\left( A^{e\alpha }f_{d[b}^{A}\frac{\partial
f_{c]e}^{B}}{\partial \varphi _{a}}+V^{E\alpha }f_{d[b}^{A}\frac{\partial
g_{c]E}^{B}}{\partial \varphi _{a}}\right) \epsilon ^{b}\xi ^{c}  \nonumber
\\
&&+\left( A^{c\alpha }\left( g_{dC}^{A}\frac{\partial f_{bc}^{B}}{\partial
\varphi _{a}}+f_{db}^{A}\frac{\partial g_{cC}^{B}}{\partial \varphi _{a}}%
\right) -V^{D\alpha }\left( g_{dC}^{A}\frac{\partial g_{bD}^{B}}{\partial
\varphi _{a}}-f_{db}^{A}\frac{\partial f_{CD}^{B}}{\partial \varphi _{a}}%
\right) \right)  \nonumber \\
&&\left. \times \left( \epsilon ^{b}\xi ^{C}-\xi ^{b}\epsilon ^{C}\right)
\right) +g\frac{\delta \bar{S}_{0}}{\delta H_{\mu }^{e}}\left( -\frac{%
\partial ^{2}W_{bc}}{\partial \varphi _{a}\partial \varphi _{e}}\left(
\epsilon ^{b}\xi _{\mu \nu }^{c}-\xi ^{b}\epsilon _{\mu \nu }^{c}\right)
\right.  \nonumber \\
&&+\left( \frac{\partial ^{3}W_{bc}}{\partial \varphi _{a}\partial \varphi
_{d}\partial \varphi _{e}}B_{d\mu \nu }-\frac{\partial ^{2}f_{bc}^{A}}{%
\partial \varphi _{a}\partial \varphi _{e}}\bar{F}_{A\mu \nu }\right)
\epsilon ^{b}\xi ^{c}  \nonumber \\
&&\left. -\bar{F}_{A\mu \nu }\left( \frac{\partial ^{2}g_{bB}^{A}}{\partial
\varphi _{a}\partial \varphi _{e}}\left( \epsilon ^{b}\xi ^{B}-\xi
^{b}\epsilon ^{B}\right) +\frac{\partial ^{2}f_{BC}^{A}}{\partial \varphi
_{a}\partial \varphi _{e}}\epsilon ^{B}\xi ^{C}\right) \right)  \nonumber \\
&&+g^{2}\delta _{\mu }^{[\alpha }\delta _{\nu }^{\beta ]}M_{\alpha }^{ab\nu }%
\frac{\delta \bar{S}_{0}}{\delta H^{b\beta }},  \label{com4}
\end{eqnarray}

\begin{eqnarray}
&&\left[ \bar{\delta}_{\epsilon },\bar{\delta}_{\xi }\right] V_{\mu }^{A}=%
\bar{\delta}_{\Lambda }V_{\mu }^{A}-g\frac{\delta \bar{S}_{0}}{\delta
H^{a\mu }}\left( \frac{\partial f_{BC}^{A}}{\partial \varphi _{a}}\epsilon
^{B}\xi ^{C}+\frac{\partial f_{bc}^{A}}{\partial \varphi _{a}}\epsilon
^{b}\xi ^{c}\right.  \nonumber \\
&&\left. +\frac{\partial g_{bB}^{A}}{\partial \varphi _{a}}\left( \epsilon
^{b}\xi ^{B}-\xi ^{b}\epsilon ^{B}\right) \right) ,  \label{com5}
\end{eqnarray}
where 
\begin{equation}
\Lambda ^{\alpha _{1}}=\left( \Lambda ^{a},\Lambda _{\mu \nu }^{a},\Lambda
^{A}\right) ,  \label{not0}
\end{equation}
with 
\begin{equation}
\Lambda ^{a}\equiv g\frac{\partial W_{bc}}{\partial \varphi _{a}}\epsilon
^{b}\xi ^{c},  \label{not1}
\end{equation}
\begin{eqnarray}
\Lambda _{\mu \nu }^{a} &\equiv &g\left( \frac{\partial f_{bc}^{A}}{\partial
\varphi _{a}}\bar{F}_{A\mu \nu }-\frac{\partial ^{2}W_{bc}}{\partial \varphi
_{a}\partial \varphi _{d}}B_{d\mu \nu }\right) \epsilon ^{b}\xi ^{c}+g\frac{%
\partial f_{BC}^{A}}{\partial \varphi _{a}}\bar{F}_{A\mu \nu }\epsilon
^{B}\xi ^{C}  \nonumber \\
&&+g\frac{\partial g_{bB}^{A}}{\partial \varphi _{a}}\bar{F}_{A\mu \nu
}\left( \epsilon ^{b}\xi ^{B}-\xi ^{b}\epsilon ^{B}\right) -g\frac{\partial
W_{bc}}{\partial \varphi _{a}}\left( \epsilon ^{b}\xi _{\mu \nu }^{c}-\xi
^{b}\epsilon _{\mu \nu }^{c}\right) ,  \label{not2}
\end{eqnarray}
and respectively 
\begin{equation}
\Lambda ^{A}\equiv g\left( f_{BC}^{A}\epsilon ^{B}\xi
^{C}+f_{ab}^{A}\epsilon ^{a}\xi ^{b}+g_{bB}^{A}\left( \epsilon ^{b}\xi
^{B}-\xi ^{b}\epsilon ^{B}\right) \right) .  \label{not3}
\end{equation}
At the same time, the function from~(\ref{com4}) denoted by $M_{\alpha
}^{ab\nu }$ reads 
\begin{eqnarray}
M_{\alpha }^{ab\nu } &=&-k_{AB}\left( \left( \left( \frac{\partial g_{cC}^{A}%
}{\partial \varphi _{b}}V_{\alpha }^{C}+\frac{\partial f_{ce}^{A}}{\partial
\varphi _{b}}A_{\alpha }^{e}\right) \left( \frac{\partial g_{dC}^{B}}{%
\partial \varphi _{a}}V^{C\nu }+\frac{\partial f_{df}^{B}}{\partial \varphi
_{a}}A^{f\nu }\right) \right. \right.  \nonumber \\
&&\left. -\left( \frac{\partial g_{dC}^{A}}{\partial \varphi _{b}}V_{\alpha
}^{C}+\frac{\partial f_{de}^{A}}{\partial \varphi _{b}}A_{\alpha
}^{e}\right) \left( \frac{\partial g_{cC}^{B}}{\partial \varphi _{a}}V^{C\nu
}+\frac{\partial f_{cf}^{B}}{\partial \varphi _{a}}A^{f\nu }\right) \right)
\epsilon ^{c}\xi ^{d}  \nonumber \\
&&+\left( \left( \frac{\partial f_{CE}^{A}}{\partial \varphi _{b}}V_{\alpha
}^{E}-\frac{\partial g_{eC}^{A}}{\partial \varphi _{b}}A_{\alpha
}^{e}\right) \left( \frac{\partial f_{DF}^{B}}{\partial \varphi _{a}}V^{F\nu
}-\frac{\partial g_{dD}^{B}}{\partial \varphi _{a}}A^{d\nu }\right) \right. 
\nonumber \\
&&\left. -\left( \frac{\partial f_{DE}^{A}}{\partial \varphi _{b}}V_{\alpha
}^{E}-\frac{\partial g_{cD}^{A}}{\partial \varphi _{b}}A_{\alpha
}^{e}\right) \left( \frac{\partial f_{CF}^{B}}{\partial \varphi _{a}}V^{F\nu
}-\frac{\partial g_{dC}^{B}}{\partial \varphi _{a}}A^{d\nu }\right) \right)
\epsilon ^{C}\xi ^{D}  \nonumber \\
&&+\left( \left( \frac{\partial g_{cD}^{A}}{\partial \varphi _{b}}V_{\alpha
}^{D}+\frac{\partial f_{ce}^{A}}{\partial \varphi _{b}}A_{\alpha
}^{e}\right) \left( \frac{\partial f_{CF}^{B}}{\partial \varphi _{a}}V^{F\nu
}-\frac{\partial g_{dC}^{B}}{\partial \varphi _{a}}A^{d\nu }\right) \right. 
\nonumber \\
&&\left. -\left( \frac{\partial f_{CE}^{A}}{\partial \varphi _{b}}V_{\alpha
}^{E}-\frac{\partial g_{eC}^{A}}{\partial \varphi _{b}}A_{\alpha
}^{e}\right) \left( \frac{\partial g_{cD}^{B}}{\partial \varphi _{a}}V^{D\nu
}+\frac{\partial f_{cd}^{B}}{\partial \varphi _{a}}A^{d\nu }\right) \right)
\times  \nonumber \\
&&\left. \times \left( \epsilon ^{c}\xi ^{C}-\xi ^{c}\epsilon ^{C}\right)
\right) .  \label{not7}
\end{eqnarray}
From the terms of antighost numbers three and four present in~(\ref{f51}) we
can recover the higher-order structure functions due to the open character
of the deformed gauge algebra, as well as the accompanying identities. They
have an intricate (but not illuminating) form and consequently we will omit
writing their concrete expressions.

At this point, we have all the information on the gauge structure of the
deformed model, whose free limit is given by the Lagrangian action~(\ref{f1}%
), together with the abelian and irreducible gauge symmetries~(\ref{f3}). We
observe that there are two main types of vertices in the deformed action (%
\ref{f53}). The first kind 
\begin{equation}
g\left( H_{\mu }^{a}W_{ab}A^{b\mu }+\frac{1}{2}B_{a}^{\mu \nu }\frac{%
\partial W_{bc}}{\partial \varphi _{a}}A_{\mu }^{b}A_{\nu }^{c}\right) ,
\label{purebf}
\end{equation}
corresponds to the self-interactions among the purely BF fields in the
absence of the vector fields $\left\{ V_{\mu }^{A}\right\} $, being given
only by terms of order one in the coupling constant. Such terms have been
previously obtained in the literature and we will not insist on their
structure (for a detailed analysis, see for instance~\cite{mpla}). The
second kind of vertices can be written in the form 
\begin{eqnarray}
&&-gk_{AB}\left( \partial ^{\mu }V^{\nu A}\right) \left( f_{CD}^{B}\left(
\varphi \right) V_{\mu }^{C}V_{\nu }^{D}+f_{ab}^{B}\left( \varphi \right)
A_{\mu }^{a}A_{\nu }^{b}+g_{aC}^{B}\left( \varphi \right) A_{[\mu
}^{a}V_{\nu ]}^{C}\right)  \nonumber \\
&&-\frac{g^{2}}{4}k_{AD}\left( f_{BC}^{A}\left( \varphi \right) V_{\mu
}^{B}V_{\nu }^{C}+f_{ab}^{A}\left( \varphi \right) A_{\mu }^{a}A_{\nu
}^{b}+g_{aB}^{A}\left( \varphi \right) A_{[\mu }^{a}V_{\nu ]}^{B}\right)
\times  \nonumber \\
&&\times \left( f_{EF}^{D}\left( \varphi \right) V^{E\mu }V^{F\nu
}+f_{cd}^{D}\left( \varphi \right) A^{c\mu }A^{d\nu }+g_{cE}^{D}\left(
\varphi \right) A^{c[\mu }V^{\nu ]E}\right) .  \label{crosscoup}
\end{eqnarray}
We note that~(\ref{crosscoup}) contains a vertex involving only the BF
fields, namely $-\frac{1}{4}k_{AD}f_{ab}^{A}\left( \varphi \right)
f_{cd}^{D}\left( \varphi \right) A_{\mu }^{a}A_{\nu }^{b}A^{c\mu }A^{d\nu }$%
, whose existence is induced by the presence of the vector fields $\left\{
V_{\mu }^{A}\right\} $. Indeed, if the vector fields $\left\{ V_{\mu
}^{A}\right\} $ were absent ($k_{AB}=0$), then this term would vanish. The
remaining terms reveal the cross-couplings between the BF fields and the
vector fields $\left\{ V_{\mu }^{A}\right\} $. Among the cross-coupling
pieces in~(\ref{crosscoup}), we find generalized cubic and quartic
Yang-Mills-like vertices in `backgrounds' of the scalar fields. We remark
that neither the one-forms $\left\{ H_{\mu }^{a}\right\} $ nor the two-forms 
$\left\{ B_{a}^{\mu \nu }\right\} $ can be coupled in a consistent,
non-trivial manner to the vector fields. Related to the deformed gauge
transformations~(\ref{f55a}--\ref{f55d}), there appears a complementary
situation, in the sense that among the BF fields, only the one-forms $%
\left\{ H_{\mu }^{a}\right\} $ and the two-forms $\left\{ B_{a}^{\mu \nu
}\right\} $ gain gauge symmetries involving the parameters $\epsilon ^{B}$.
The deformed gauge transformations of the vector fields $\left\{ V_{\mu
}^{A}\right\} $ have a rich structure, including, besides other terms, the
generalized covariant derivative $\left( \delta _{\;\;B}^{A}\partial _{\mu
}-gf_{BC}^{A}\left( \varphi \right) V_{\mu }^{C}\right) $ with respect to
the parameters $\epsilon ^{B}$.

\section{Some solutions to the Eqs.~(\ref{relcoef})}

We have seen that the deformation procedure developed so far essentially
relies on the existence of four types of functions depending on the
undifferentiated scalar fields, namely, $W_{ab}$, $f_{BC}^{A}$, $f_{ab}^{A}$
and $g_{aB}^{A}$, which are subject on the one hand to the conditions~(\ref
{fas}) plus~(\ref{f31}) and on the other hand to the Eqs.~(\ref{relcoef}%
). In the sequel we analyze two classes of solutions to the above conditions
and equations and emphasize that they admit an interesting geometric and
algebraic interpretation. The first class of solutions corresponds to a
non-vanishing $W_{ab}\left( \varphi \right) $, while the second kind is
associated with $W_{ab}=0$.

Related to the former type of solutions, it is clear that the first equation
from~(\ref{relcoef}) together with the first antisymmetry property in~(\ref
{fas}) 
\begin{equation}
W_{e[a}\frac{\partial W_{bc]}}{\partial \varphi _{e}}=0,\;W_{ab}=-W_{ba},
\label{f43a}
\end{equation}
shows that the antisymmetric functions $W_{ab}$ of the undifferentiated
scalar fields satisfy the Jacobi's identity for a nonlinear algebra. Let us
see the geometric meaning of $W_{ab}$. To this end, we briefly review the
basic notions on Poisson manifolds. If $N$ denotes an arbitrary Poisson
manifold, then this is equipped with a Poisson bracket $\left\{ ,\right\} $
that is bilinear, antisymmetric, subject to a Leibnitz-like rule and
satisfies a Jacobi-type identity. If $\left\{ X^{i}\right\} $ are some local
coordinates on $N$, then there exists a two-tensor $\mathcal{P}^{ij}\equiv
\left\{ X^{i},X^{j}\right\} $ (the Poisson tensor) that uniquely determines
the Poisson structure together with the Leibnitz rule. This two-tensor is
antisymmetric and transforms covariantly under coordinate transformations.
Jacobi's identity for the Poisson bracket $\left\{ ,\right\} $ expressed in
terms of the Poisson tensor reads as $\mathcal{P}_{,k}^{ij}\mathcal{P}^{kl}+%
\mathrm{cyclic}\left( i,j,l\right) =0$, where $\mathcal{P}_{,k}^{ij}\equiv
\partial \mathcal{P}^{ij}/\partial X^{k}$. Now, the geometric origin of $%
W_{ab}$ is obvious. If, for instance, we choose a concrete form for the
antisymmetric functions $W_{ab}\left( \varphi \right) $ that satisfy~(\ref
{f43a}), then we can interpret the dynamical scalar fields $\left\{ \varphi
_{a}\right\} $ precisely like some local coordinates on a target manifold
endowed with a prescribed Poisson structure (up to the plain convention that
the lower index $a$ is a `covariant' index of the type $i$). Conversely, any
given Poisson manifold parametrized in terms of some local coordinates $%
\left\{ \varphi _{a}\right\} $ (within the same index convention) prescribes
a Poisson tensor $W_{ab}\left( \varphi \right) $ which is antisymmetric and
satisfies~(\ref{f43a}). Once we have fixed the functions $W_{ab}$, it is
easy to see that a solution for the remaining coefficients ($f_{BC}^{A}$, $%
f_{ab}^{A}$ and $g_{aB}^{A}$) is represented by 
\begin{equation}
f_{BC}^{A}=\bar{f}_{BC}^{A},\;g_{aB}^{A}=\bar{f}_{BE}^{A}M^{E}X_{a}\left(
\varphi _{c}\right) ,  \label{f44a}
\end{equation}
\begin{equation}
f_{ab}^{A}=M^{A}\left( X_{c}\frac{\partial W_{ab}}{\partial \varphi _{c}}%
+W_{ca}\frac{\partial X_{b}}{\partial \varphi _{c}}+W_{bc}\frac{\partial
X_{a}}{\partial \varphi _{c}}\right) ,  \label{f44b}
\end{equation}
where $M^{E}$ are some real constants, $\left\{ X_{a}\right\} $ stands for a
set of arbitrary functions depending only on the undifferentiated scalar
fields $\varphi _{a}$, and $\bar{f}_{BC}^{A}$ are some real, antisymmetric
constants, that obey the identity 
\begin{equation}
\bar{f}_{E[B}^{A}\bar{f}_{CD]}^{E}=0.  \label{f45}
\end{equation}
Accordingly, $\bar{f}_{BC}^{A}$ can be viewed like the structure constants
of a semi-simple Lie algebra, endowed with the Killing-Cartan metric $k_{AB}$%
, while $M^{A}$ can be seen like the components of an arbitrary element from
this Lie algebra. In this situation, the deformed Lagrangian action~(\ref
{f53}) also includes self-interactions among the vector fields $\left\{
V_{\mu }^{A}\right\} $ precisely described by cubic and quartic Yang-Mills
vertices. Accordingly, the gauge transformations of $V_{\mu }^{A}$ contain
the well-known covariant derivative of the gauge parameters $\epsilon ^{A}$%
\begin{equation}
\bar{\delta}_{\epsilon }V_{\mu }^{A}=\left( \delta _{\;\;B}^{A}\partial
_{\mu }-g\bar{f}_{BC}^{A}V_{\mu }^{C}\right) \epsilon ^{B}+``\mathrm{more}".
\label{f45abcdef}
\end{equation}

Next, we examine the latter kind of solutions (corresponding to $W_{ab}=0$),
in which case the Eqs.~(\ref{relcoef}) become 
\begin{equation}
f_{E[B}^{A}f_{CD]}^{E}=0,  \label{relc1}
\end{equation}
\begin{equation}
f_{[ab}^{E}g_{c]E}^{A}=0,%
\;f_{EB}^{A}f_{ab}^{E}+g_{aB}^{E}g_{bE}^{A}-g_{bB}^{E}g_{aE}^{A}=0,
\label{relcn}
\end{equation}
\begin{equation}
g_{aE}^{A}f_{BC}^{E}+f_{EB}^{A}g_{aC}^{E}-f_{EC}^{A}g_{aB}^{E}=0.
\label{relc3}
\end{equation}
The solution to~(\ref{relcn}--\ref{relc3}) takes the form 
\begin{equation}
f_{ab}^{A}=f_{BC}^{A}M_{a}^{B}M_{b}^{C},\;g_{aB}^{A}=f_{CB}^{A}M_{a}^{C},
\label{f56}
\end{equation}
where $M_{b}^{B}$ are some arbitrary functions of the undifferentiated
scalar fields and $f_{BC}^{A}$ verify the Eq.~(\ref{relc1}). In order
to solve the remaining equation, namely,~(\ref{relc1}), let $\bar{M}_{A}^{a}$
be some functions of the scalar fields such that $\bar{M}_{A}^{a}M_{b}^{A}=%
\delta _{b}^{a}$. Then, the solution of~(\ref{relc1}) reads as 
\begin{equation}
f_{BC}^{A}=f_{bc}^{a}M_{a}^{A}\bar{M}_{B}^{b}\bar{M}_{C}^{c},  \label{f57}
\end{equation}
where $f_{bc}^{a}$ are the structure constants of a semi-simple Lie algebra
with the Killing-Cartan metric $k_{ab}$. It is easy to see that the Jacobi
identity~(\ref{relc1}) is a consequence of the Jacobi identity for the
structure constants $f_{bc}^{a}$. For the functions $g_{aB}^{A}$ and $%
f_{BC}^{A}$ given in~(\ref{f56}--\ref{f57}) to satisfy the antisymmetry
properties~(\ref{fas}) and~(\ref{f31}), it is necessary that $k_{AB}$ and $%
k_{ab}$ are correlated through some relations of the type 
\begin{equation}
k_{AB}M_{a}^{A}M_{b}^{B}=k_{ab}\Phi \left( \varphi \right) ,  \label{f58}
\end{equation}
with $\Phi \left( \varphi \right) $ a non-vanishing, but otherwise arbitrary
function of the scalar fields. We remark that, although the functions $%
f_{BC}^{A}$ from~(\ref{f57}) depend in general on the scalar fields, they
however verify the Jacobi identity~(\ref{relc1}). Accordingly, these
functions can be regarded like some `structure constants' of a Lie algebra
whose generators depend on the scalar fields (generalized Lie algebra). It
is interesting to note that the gauge algebra is open also for the latter
kind of solutions. In both cases, the entire gauge structure of the
interacting model can be obtained by substituting the solutions~(\ref{f44a}--%
\ref{f44b}) and respectively~(\ref{f56}--\ref{f57}) in the formulas~(\ref
{f53}) and~(\ref{f55a}--\ref{com5}).

\section{Conclusion}

To conclude with, in this paper we have investigated the consistent
two-dimensional interactions that can be introduced among a set of scalar
fields, two types of one-forms, a system of two-forms and a collection of
vector fields, described in the free limit by an abelian BF theory and a sum
of Maxwell actions. Starting with the BRST differential for the free theory, 
$s=\delta +\gamma $, we compute the consistent first-order deformation of
the solution to the master equation with the help of some cohomological
techniques, and obtain that it is parametrized by five kinds of functions
depending on the undifferentiated scalar fields. Next, we investigate the
second-order deformation, whose existence reduces the number of independent
types of functions on the scalar fields to four and, meanwhile, requires
that these are subject to certain equations. Based on these restrictions, we
determine the expression of the second-order deformation and, moreover, show
that we can take all the remaining higher-order deformations to vanish. As a
consequence of our procedure, we are led to an interacting gauge theory with
deformed gauge transformations and a non-abelian gauge algebra that only
closes on-shell. The presence of the collection of vector fields brings in a
rich structure of non-trivial terms if compared with the self-interactions
that can be added to a two-dimensional abelian BF theory~\cite{mpla}.
Finally, we give two classes of solutions to the equations satisfied by the
various functions of the scalar fields that parametrize the deformed
solution to the master equation, which can be interpreted in terms of
Poisson manifolds and respectively of generalized Lie algebras.

\section*{Acknowledgment}

The authors wish to thank C. Bizdadea and S. O. Saliu for useful discussions
and comments. This work has been supported by the type $\mathrm{A}_{\mathrm{T%
}}$ grant 33547/2003, code 302/2003, from the Romanian Council for Academic
Scientific Research (CNCSIS) and the Romanian Ministry of Education,
Research and Youth (MECT).

\end{document}